\documentclass[12pt,a4paper]{article}
\usepackage{epsfig}

\begin{document}
\begin{flushright}AEI-2009-105\\PUPT-2319
\end{flushright}
\begin{center}

\vspace{1cm}

{\bf \large CONSTRUCTION OF INFRARED FINITE OBSERVABLES\\[0.3cm] IN  ${\cal N}=4$ SUPER
YANG-MILLS THEORY} \vspace{2cm}

{\bf \large L. V. Bork$^{2}$, D. I. Kazakov$^{1,2}$, G. S. Vartanov$^{1,3}$, \\[0.2cm]
and A. V. Zhiboedov$^{1,4}$}\vspace{0.5cm}

{\it $^1$Bogoliubov Laboratory of Theoretical Physics, Joint
Institute for Nuclear Research, Dubna, Russia, \\
$^2$Institute for Theoretical and Experimental Physics, Moscow, Russia, \\
$^3$Max-Planck-Institut f\"ur Gravitationsphysik, Albert-Einstein-Institut 14476 Golm,
Germany,\\ $^4$Department of Physics, Princeton University, Princeton, NJ 08544,
USA.}\vspace{1cm}

\abstract{In this paper we give all the details of the calculation that we presented in
our previous paper ArXiv:0908.0387 where the infrared structure of the MHV gluon
amplitudes in the planar limit for ${\cal N}=4$ super Yang-Mills theory was considered in
the next-to-leading order of perturbation theory.  Explicit cancellation  of the infrared
divergencies in properly defined inclusive cross-sections is demonstrated first in a toy
model example of "conformal QED" and then in the real ${\cal N}=4$ SYM theory. We give
the full-length details both for the calculation of the real emission and for the
diagrams with splitting in initial and final states. The finite parts for some inclusive
differential cross-sections are presented in an analytical form. In general, contrary to
the virtual corrections, they do not reveal any simple structure. An example of the
finite part containing just the log functions is presented. The dependence  of inclusive
cross-section on the external scale related to the definition of asymptotic states is
discussed.}
\end{center}


violating amplitudes

PACS classification codes: 11.15.-q; 11.30.Pb; 11.25.Tq

\newpage

\tableofcontents{}\vspace{0.5cm}

{\bf References} \hfill {\bf 45}

\newpage
\renewcommand{\theequation}{\thesection.\arabic{equation}}
\section{Introduction}\label{s1}

In recent years remarkable progress in understanding the structure of the planar
limit\footnote{Defined as $g \rightarrow 0$; $N_c \rightarrow \infty$; $\lambda = g^2 N_c
~ fixed$ } of the ${\cal N}=4$ SYM (supersymmetric Yang-Mills) theory has been achieved.
In the planar limit this theory seems to be integrable at the quantum level and its
possible solution would be the first example of a solvable nontrivial four-dimensional
Quantum Field Theory. The objects which were in the spotlight starting from the AdS/CFT
(Anti de Sitter/Con\-for\-mal Field Theory) correspondence \cite{AdsM} were the local
operators, namely,  the spectrum of their anomalous dimensions. They were calculated on
the one hand side from the field theory approach \cite{4-loop} and, on the other hand, as
energy levels of a string in the classical background \cite{GPK,FroTse03} revealing a
remarkable coincidence. This coincidence being part of the general conjecture suggests
the way towards solution of the model at the quantum level.

\subsection{Scattering amplitudes at weak coupling}
Other quantities of interest are the so-called MHV\footnote{MHV (maximally helicity
violating) amplitudes are the amplitudes where all particles are treated as outgoing and
the net helicity is equal to $n-4$ where $n$ is the number of particles. For gluon
amplitudes MHV amplitudes are defined as the amplitudes in which all but two gluons have
positive helicities.} scattering amplitudes.  It was realized long ago that in the planar
limit the pure non-Abelian gauge theories they do have a truly simple
structure~\cite{Parke}. In papers \cite{Bern1,Bern2,Bern3}, the powerful tool for
calculating the loop expansion for these amplitudes was suggested which allows one to
calculate the loop contributions to the amplitudes without calculating the usual Feynman
diagrams the number of which grows exponentially with the growth of the order of
perturbation theory. Even greater simplification occurs in the case of the ${\cal N}=4$
SYM theory where the loop expansion takes extremely simple form in comparison with a less
supersymmetric case \cite{Green}.

To see the hidden symmetries of the MHV amplitudes, it
is useful to consider the color-ordered amplitude defined through
the group structure decomposition
\begin{eqnarray}
\mathcal{A}_{n}^{(l-loop)}=g^{n-2} \left(\frac{g^2N_c}{16\pi^2}
\right)^{l}\sum_{perm}
Tr(T^{a_{\rho(1)}}...T^{a_{\rho(n)}})A^{(l)}_n(p_{\rho(1)},...,p_{\rho(n)}),
\end{eqnarray}
where $\mathcal{A}_{n}$ is the physical amplitude, $A_n$ are the
partial color-ordered amplitudes, $T^{a(i)}$ are the generators of
the gauge group $SU(N_c)$, $a_{\rho(i)}$ is the color index of the
$\rho(i)$-th external particle, and $p_{\rho(i)}$ is its momentum.

To be more precise, it was found that these amplitudes revealed  the iterative structure
which was first established in two loops~\cite{Anastasiou:2003kj} and then confirmed at
the three loop level by Bern, Dixon and Smirnov, who formulated the ansatz \cite{bds05}
for the all-loop $n$-point MHV amplitudes:
\begin{equation}
{\cal M}_n \equiv \frac{A_{n}}{A_{n}^{tree}}\! =\!
1+\sum\limits_{L=1}^\infty\left(\frac{g^2N_c}{16\pi^2}\right)^L\!\!\! M_n^{(L)}(\epsilon)
=\exp\!\left[\sum\limits_{l=1}^\infty\!\left(\!\frac{g^2N_c}{16\pi^2}\right)^l\!\!\!
\left(f^{(l)}(\epsilon)
M_n^{(1)}(l\epsilon)\!+\!C^{(l)}\!+\!E^{(l)}_n(\epsilon)\right)\right],
\end{equation}
where $E_n^{(l)}$  vanishes as $\epsilon \rightarrow 0$, $C^{(l)}$
are some finite constants, and $M_n^{(1)}(l\epsilon)$ is the
$l\epsilon$- regulated one-loop $n$-point amplitude.

It is not surprising that the IR divergent parts of the amplitudes
factorize and exponentiate \cite{infrared}. What is less obvious  is
that it is also true for the finite part
\begin{eqnarray}
{\cal M}_n(\epsilon)&=&\exp\left[-\frac
18\sum\limits_{l=1}^\infty\left(\frac{g^2N_c}{16\pi^2}\right)^l
\left(\frac{\gamma^{(l)}_{cusp}}{(l\epsilon)^2}+\frac{2G^{(l)}_0}{l\epsilon}\right)
\sum\limits_{i=1}^n \left(\frac{\mu^2}{-s_{i,i+1}}\right)^{l\epsilon}\right.\nonumber\\
&&\left.+\frac 14
\sum\limits_{l=1}^\infty\left(\frac{g^2N_c}{16\pi^2}\right)^l
\gamma^{(l)}_{cusp} F^{(1)}_n(0)+C(g) \right],
\end{eqnarray}
where $\gamma_{cusp}(g) = \sum_l (\frac{g^2N_c}{16\pi^2})^l
\gamma^{(l)}_{cusp}$ is the so-called cusp anomalous dimension
\cite{colK} and $G_0(g) = \sum_l(\frac{g^2N_c}{16\pi^2})^l G_0^{(l)}$ is the second
function (dependent on  the IR regularization) which defines the IR
structure of the amplitude.

According to the BDS ansatz, the finite part of the amplitude is defined by the cusp
anomalous dimension and a function of kinematic parameters specified at one-loop. For a
four gluon amplitude one has
\begin{equation}\label{fin4}
  F^{(1)}_4(0)= \frac{1}{2} \log^2 \left( \frac{-t}{s} \right) + 4 \zeta_2.
  \end{equation}

 The cusp anomalous dimension is a function of the gauge coupling, for which four terms of
the weak coupling expansion~\cite{4-loop} and three terms of the
strong coupling expansion~\cite{GPK,FroTse03,RoibanTseytlin} are
known. Integrability from the both sides of the AdS/CFT
correspondence leads  to the all-order integral
equation~\cite{BES06} solution to which, being expanded in the
coupling, reproduces both series~\cite{BaKo}.

For $n=4,5$ the BDS ansatz goes through all checks, namely, the amplitudes were
calculated up to four loops for four gluons \cite{4-loop} for
divergent terms (see also \cite{SpV} for checking at order
$1/\epsilon$) up to two loops for five gluons \cite{chazo} and up to
three loops in \cite{SpVW}. However, starting from $n=6$ it fails. The first indication of the
problem was the strong coupling calculation in the limit $n
\rightarrow \infty$ \cite{am2} where the authors compute the value
of the amplitude for a particular kinematic configuration for a
large number of gluons and find that the result disagrees with the
exact value of the amplitude from the BDS formula. The second
indication came also from this duality, namely, from the comparison
of hexagonal light-like Wilson loop and finite part of the BDS
ansatz for the six-gluon amplitude. It was found that the two
expressions differ by a nontrivial function of the three (dual)
conformally invariant variables \cite{Kor2l}. The third indication
appeared in  \cite{Lip} where the analytical structure of the BDS
ansatz was analyzed and starting from $n=6$ the Regge limit
factorization of the amplitude in  some physical regions failed.
Finally, it was shown by explicit two-loop calculation
\cite{twoBern} that the BDS ansatz is not true and it needs to be
modified by some unknown finite function, which  is an open and
intriguing problem. However, from the two-loop calculation for the
six-point amplitude \cite{twoBern} and hexagonal light-like Wilson
loop \cite{KorVer} it was shown that the gluon amplitude/Wilson loop
duality \cite{Dks} is still valid.

\subsection{Strong coupling dual of amplitudes, light-like Wilson loops
and dual conformal invariance}

In \cite{am1}, the authors defined the prescription for calculating the amplitudes at
strong coupling.  It happens that in leading order the amplitude is given by the
light-like Wilson loop living on the boundary of dual $AdS$ space
\begin{eqnarray}\label{WLAds}
{\cal M}_n \sim \exp [ - S_{cl}^{E}] = \exp
[\frac{\sqrt{\lambda}}{2\pi} (Area)_{cl}],
\end{eqnarray}
where $S_{cl}^E$ denotes  the classical action of  classical solution of the string
worldsheet equations in Euclidean space-time, which is proportional to the area of the
string world-sheet.

After this  in \cite{Dks} it was conjectured that duality between light-like Wilson loops
and MHV scattering amplitudes is valid at any coupling, which was proved for $n$-point
MHV amplitudes at one loop \cite{Brand} and for $n=6$ at two loops \cite{KorVer} (for
more details and  references see the review \cite{AldayRoiban}).

Due to the cusps the light-like Wilson  loop is UV divergent; however, this divergency is
under control, namely one can write the divergent factor in all orders in the coupling
governed by two functions,  one of them being  the cusp anomalous dimension mentioned
above. This allows one to define the finite parts for both the Wilson loop with $n$ cusps
and the $n$--point MHV amplitude which, according to DKS conjecture~\cite{Dks}, are equal
to each other
\begin{eqnarray}
Fin [\log {\cal M}_n] = Fin [\log {\cal W}_n].
\end{eqnarray}

In \cite{Dhks1} the notion of dual superconformal symmetry was introduced, which is
conformal invariance acting in momentum space. What is important, this symmetry has a
non-Lagrangian nature. After this in \cite{Berkovits:2008ic, RT} the fermionic
$T$-duality was suggested which maps the dual superconformal symmetry of the original
theory to the ordinary superconformal symmetry of the dual model.

For a Wilson loop the conformal invariance is broken due to the cusps, but one can write
the anomalous Ward identities which allows one to find the finite parts of the Wilson
loop with $n=4$ and $n=5$ cusps exactly \cite{Dhks2}
\begin{eqnarray}
\sum_{i=1}^{n}(2x_i^{\nu}x_i \partial_i -x_i^2 \partial_i^{\nu}) Fin [\log {\cal W}_n] =
\frac{1}{2}\gamma_{cusp} \sum_{i=1}^n \log \frac{x_{i,i+2}^2}{x_{i-1,i+1}^2}
x_{i,i+1}^{\nu},
\end{eqnarray}
where the connection between the momentum space and its dual
$x_{i,i+1}^{\mu}=x_i^{\mu}-x_{i+1}^{\mu}=p^{\mu}_i$ is used. This equation uniquely fixes
the finite parts of the Wilson loop with $n=4$ and $n=5$ cusps; however, starting from
$n=6$ more input is needed since the finite part of the Wilson loop in this case can be a
function of the three conformal invariant variables. Hopefully, one can find hidden
symmetries which fix the finite part for any $n$ \cite{yangian}.

It is not clear how to derive this duality from the field theory point of view, and also
how to extend it to the NMHV case\footnote{NMHV (next to maximally helicity violating)
amplitudes are the amplitudes where all particles are treated as outgoing and the net
helicity is equal to $n-6$ where $n$ is the number of particles.}. At one-loop one can
show that finite part of the so-called two mass easy box which governs the finite
function of MHV amplitudes could be directly mapped to Wilson loop diagrams through a
simple change of variables in the space of Feynman parameters and also through the
connection between scalar integrals in different dimensions \cite{Gorsky:2009nv}.

\subsection{Infrared-safe observables}

While all the UV divergences in ${\cal N}=4$ SYM are absent in scattering amplitudes the
IR ones remain and are supposed to be canceled in  properly defined quantities. By
themselves the divergent amplitudes have no sense. Regularized expressions act like some
kind of scaffolding which has to be removed to obtain eventual physical observables. It
is these quantities that are the aim of our calculation. And though the
Kinoshita-Lee-Nauenberg~\cite{KLN} theorem in principle tells us how to construct such
quantities, explicit realization of this procedure is not simple and one can think of
various possibilities. The well known example is a successful application to observables
in QED \cite{Weinberg}. The other suggestion is to consider the so-called energy flow
functions defined in terms of the energy-momentum tensor correlators
introduced earlier (see for example \cite{BBEL})  and considered in the
weak coupling regime in \cite{EF,EFK} and recently in the strong
coupling regime in \cite{hofmal}.  From our side we concentrated on inclusive
cross-sections in hope that they reveal some factorization properties discovered in the
regularized amplitudes. Similar questions were discussed in \cite{vanNeerven:1985ja},
where the inclusive cross-sections  like the IR safe observables based on on-shell
formfactors in $\cal{N}$=4 SYM were constructed.

To perform the procedure of cancellation of the IR divergences, one should have in mind
that in conformal theory all the masses are zero and one has additional collinear
divergences which need special care. In this work we employ the method developed in the
QCD parton model \cite{EKSglu,EKS,KunsztSoper,Kunszt3Jet,Katani}. It includes two main
ingredients in the cancellation of infrared divergencies coming from the loops:  emission
of additional soft real quanta and redefinition of the asymptotic states resulting in the
splitting terms governed by the kernels of the DGLAP equations~\cite{GrLip,Alt}. The
latter ones take care of the collinear divergences.

Typical observables in QCD parton model calculations are inclusive jet cross-sections,
where  the total energy of scattered partons is not fixed since they are considered to be
parts of the scattered hadrons. In~\cite{EKS}, the algorithm for extracting divergences
was developed which allows one to cancel divergences and apply numerical methods for
calculation of the finite part. In our paper we choose as our observables the inclusive
cross-sections  with fixed initial energy and get an analytical expression for the finite
part of the differential cross-section. We do not assume any confinement and consider the
scattering of the single parton based ``coherent" states,\footnote{the squared
perturbative amplitudes used in our calculation are summed over colors, so in this sense
they are colorless and there are no contradiction with statements that cancellation of IR
divergences occurs only for colorless objects.} being the asymptotic states of conformal
field theory.

There are some attempts to deal with the divergences for the amplitudes themselves. For
example in \cite{BT}  a deformation of the free superconformal representation by
contributions which change the number of external legs was proposed which looks similar
to the procedure that we apply below considering the  inclusive cross-sections. Acting
along the same lines the authors of \cite{SV} have observed that the holomorphic anomaly
\cite{CSW} gives an extra modification of superconformal algebra for the tree level
scattering amplitudes. They argued that superconformal symmetry survives regularization
and introduced a new holomorphic anomaly friendly regularization  to deal with the IR
divergences.

The paper is organized as follows. In Section \ref{s2}, we consider general issues
concerning the construction of the infrared-finite observables in the massless QFT. We
discuss the IR and collinear divergencies for the scattering amplitudes and the ways of
 their cancellation based on the Kinoshita-Lee-Nauenberg theorem. We introduce the notion of the
measurement functions and discuss their properties. Then the concept of the splitting
functions and splitting counterterms is outlined. We define the IR finite inclusive
cross-sections which are the subject of calculations in the subsequent sections.

Section \ref{s3} is devoted to the demonstration of the techniques discussed above in
practice. In a toy model of ``conformal QED" we consider the $\alpha_{s}$ correction to
the massless electron-quark scattering. We show how the IR and collinear divergences
cancel and calculate analytically the remaining finite part of the differential
cross-section. Due to absence of the identical particles in the final state this example
turns out to be much simpler than gluon scattering in ${\cal N}=4$ SYM and serves as a
good warm-up exercise before going to ${\cal N}=4$ SYM.

In Section \ref{s4}, we calculate the leading order PT (Perturbation
Theory) correction to the gluon-gluon scattering inclusive
cross-section. It includes the one-loop contribution to the $2\to 2$
scattering differential cross-section, the tree level $2\to 3$
scattering with the integration over the phase space of the fifth
gluon and an account of the splitting of the initial and final
states. We consider also the amplitudes with creation of pairs of
the matter fields from the ${\cal N}=4$ supermultiplet.

In Section \ref{s5}, using the results of Section \ref{s4} we present the infrared finite
results for the  differential  inclusive cross-section in ${\cal N}=4$ SYM theory for
different physical setups.

Section \ref{s6} contains discussion and concluding remarks.

In appendices we present the technical details of our calculations.

\section{Construction of Infra-Red Safe Observables}\label{s2}

The tree level matrix elements are finite and well defined in perturbation theory.
Divergences appear when integrating over virtual loops or over phase space of real
particles. So the first step is to choose a proper quantity which is finite in the lowest
order of PT prior to calculation of radiative corrections. For example, the total elastic
$2\times 2$  cross-section is divergent, but differential cross-section is well defined.
The choice of a proper quantity is performed by imposing conditions on the phase space.
This can be achieved by introducing the concept of the \textit{measurement function}
${\cal S}_n$, where $n$ is the number of particles in the final state. It defines which
physical quantity we are measuring. Typical examples are: a total cross-section, a
differential cross-section, an $n$-jet cross-section, etc. In the case of $2\times n$
scattering the differential cross-section is given by
\begin{eqnarray}
\frac{d\sigma_{2 \rightarrow n}}{d \Omega} &=& \frac{1}{J}\int
|M_{2+n}|^2d\phi_{n}{\cal S}_n,\nonumber
\end{eqnarray}
where ${\cal S}_n$ is the measurement function and the n-particle
phase space $d\phi_{n}$ is given by
\begin{eqnarray}
d\phi_{n}&=&\prod_{k=3}^{n+2}
\delta^{+}(p^2_k)\frac{d^{D}p_k}{(2\pi)^{D-1}}(2\pi)^D\delta^{D}(p_1+p_2-p_3-...-p_{n+2}).
\end{eqnarray}
Here $J$ is the flux factor, $p_1,p_2$ are the momenta of the
incoming particles, $p_3,...,p_{n+2}$ are the momenta of the
outgoing ones, $M_{2+n}$ is the matrix element of the corresponding
process and  we use the dimensional regularization with
$D=4-2\epsilon$.

Then, for example, choosing the measurement function to be
$$\mathcal{S}_2=\delta^{D-2}
(\Omega_{Det} - \Omega_{13}) ,$$ one singles out the standard
differential cross-section for the scattering of a third particle on
a certain solid angle $\Omega_{13}$ for the $2 \rightarrow 2$
process
\begin{equation}\label{cr}
  \frac{d\sigma_{2 \rightarrow
2}}{d\Omega_{13}}=\int |M_4|^2d\phi_{2}\ \mathcal{S}_2.
\end{equation}

If one wants to construct the IR finite quantity then, according  to the
Kinoshita-Lee-Nauenberg~\cite{KLN} theorem, it is not sufficient to consider the process
with the fixed number of final particles. One has to include the processes of the same
order of the perturbation theory with emission of extra  soft quanta and integrate over
their momenta. This leads  to the notion of inclusive cross-section when one fixes some
particles and integrate over all the others allowed by the conservation laws.

When the number of particles increases, one has to specify the measurable quantity in a
more accurate way and to distinguish the particle(s) in the final state. Thus, one can
introduce the energy and angular resolution for the detector and  cut the phase space so
that the soft quanta with total energy below the threshold as well as all the particles
within the given solid angle are included. This procedure requires the corresponding
measurement functions and works well in QED but introduces explicit dependence on the
energy and angular cutoff, thus violating conformal invariance.

We adopt here different attitude without introducing any cutoffs but rather considering
the inclusive cross-section with the emission of particles with all possible momenta
allowed by kinematics. Having identical particles in the final state one has to specify
which particles are detected by introducing some  measurement function. For instance, one
can detect the given particle scattering on a given angle while integrating over  the
phase space of the other particles.  As it will be clear later in this case due to
collinear divergences one still cannot avoid introducing some scale related to the
definition of the asymptotic states of a theory. Below we show how it works in particular
examples.

To have the cancellation of all the IR divergences, according to the analysis of Ellis,
Kunszt and Soper \cite{EKS}, the measurement functions for the processes with a different
number of external particles have to obey the following conditions:
\begin{eqnarray}\label{appropr2}
  {\cal S}_{n+1}(..., \lambda \vec p,...) = {\cal S}_{n}(......), \ \ \lambda\to 0,
\end{eqnarray}
which reflects the insensitivity to the soft quanta, and
  \begin{eqnarray} \label{appropr}
  {\cal S}_{n+1}(..., \lambda \vec p,..., (1 - \lambda) \vec p,...) =
  {\cal S}_{n}(...,\vec p,...)
\end{eqnarray}
here $0 \leq \lambda \leq 1$. This condition expresses our insensibility to collinear
quanta.

It should be pointed out that in case of identical particles one has an additional
problem when calculating the differential cross-sections: one has to specify the
scattering angles and to choose the detectable particle. This requirement imposes further
conditions on the phase space as will be shown below when considering the gluon
scattering.

The additional divergences appearing in the massless case which come from the integration
over angles rather than the modulus of momentum, as in the case of the IR divergences,
are related to the collinearity of momenta of two particles. For this reason they are
called {\it the collinear divergences}. To get the cancellation of all divergences, the
observed cross-section should include besides the main process and emission of the soft
quanta the process of emission of collinear particles with kinematically allowed absolute
values of momenta.  As we will see below, the leading IR divergences coming from the
cross-section of the processes with the virtual loop correction and from the real
emission of the soft quanta cancel. However, the total cancellation of divergencies does
not happen. The remaining divergences in the form of a single pole have the collinear
nature. For the cancellation of the remaining pole one has to properly define the initial
(and final) states. The reason is that a massless particle can emit a collinear one which
carries part of the initial momentum and in this case, it is impossible to distinguish
one particle propagating with the speed of light from the two flying parallel. This  is
the common problem for any theory containing the interacting massless particles.

To deal with this problem, let us consider a particle in the initial state and introduce
the notion of distribution of a particle with respect to the fraction of the carried
momentum $z$: $q(z)$. Then the zero-order distribution corresponds to $q(z)=\delta(1-z)$
and the emission of a collinear particle leads to the splitting: the particle $i$ carries
the fraction of momentum equal to $z$, while the collinear particle $j$ - $(1-z)$. The
probability of this event is given by the so-called {\it splitting functions}
$P_{ij}(z)$~\cite{Alt}. In case of a particle in a final state, this corresponds to the
fragmentation
 into a pair of particles $i$ and $j$. In the lowest order of perturbation theory the
distribution can be written in the form
\begin{eqnarray}\label{q}
q_i(z,\frac{Q_{f}^2}{\mu^2})&=&\delta(1-z)+\frac{\alpha}{2\pi}\frac{1}{\epsilon}
\left(\frac{\mu^2}{Q_{f}^2}\right)^\epsilon \sum_j P_{ij}(z),
\end{eqnarray}
where the scale $Q_{f}^2$, sometimes called factorization scale, defines  the measure of
collinearity of the emitted particles, i.e., it refers to the definition of the initial
state. In fact, in the massless case one cannot define the initial state that contains
just one particle, it exists together with the set of collinear particles forming a
coherent state.

This leads to the additional terms in the cross-section
\begin{equation}\label{initsp}
\frac{\alpha}{2\pi}\frac{1}{\epsilon} \left(\frac{\mu^2}{Q_{f}^2}\right)^\epsilon\int_0^1
dz \sum_j P_{ij}(z) \ d\sigma^{Born}_j(zp_1,p_2,p_3,p_4) + (p_1 \leftrightarrow p_2),
\end{equation}
referred hereafter as the initial splitting contributions or collinear counterterms.

The same is true for the final states. The corresponding final state collinear
counterterms are
\begin{equation}\label{finsp}
\frac{\alpha}{2\pi}\frac{1}{\epsilon} \left(\frac{\mu^2}{Q_{f}^2}\right)^\epsilon\int_0^1
dx \sum_j P_{ij}(x) \ d\sigma^{Born}_j(p_1,p_2,p_3,p_4) + (p_3 \leftrightarrow p_4).
\end{equation}

Summarizing all the contributions we come to the following set of IR safe observables
that we consider here
 \begin{eqnarray}\label{IRsafe}
d\sigma^{incl}_{obs}&=&\sum\limits_{n=2}^{\infty}\int\limits_0^1\!dz_1\
q_1(z_1, \frac{Q_f^2}{\mu^2})\! \int\limits_0^1\!dz_2\ q_2(z_2,
\frac{Q_f^2}{\mu^2})\prod\limits_{i=3}^{n+2}\int\limits_0^1\!dz_i\
q_i(z_i,
\frac{Q_f^2}{\mu^2})\times \\
   &&\hspace{-1.7cm}\times \ d\sigma^{2\to n}(z_1p_1,z_2p_2,...)S_n(\{z\})=
   g^4N_c^4\sum\limits_{L=0}^\infty \ \left(\frac{g^2N_c}{16\pi^2}\right)^L d\sigma_L^{Finite}
   (s,t,u,Q_f^2),\nonumber
\end{eqnarray}
where $p_{1},p_{2}$ are the momenta of the initial particles, $p_{i}$ are the momenta of
the final particles, ${\cal S}_{n}$ are the measurement functions which define the
measurable quantity, $q_{i}$ are the initial and final state distributions.

 The above expression looks like  the parton model cross-section. The difference is that
in the parton model one uses both the parton distributions inside
hadrons and fragmentation functions for final-state hadrons while
here it belongs to the definition of the asymptotic states.

\section{Toy model:``Conformal QED"}\label{s3}

\begin{figure}[ht]\vspace{0.2cm}
 \begin{center}
 \leavevmode
  \epsfxsize=15cm
 \epsffile{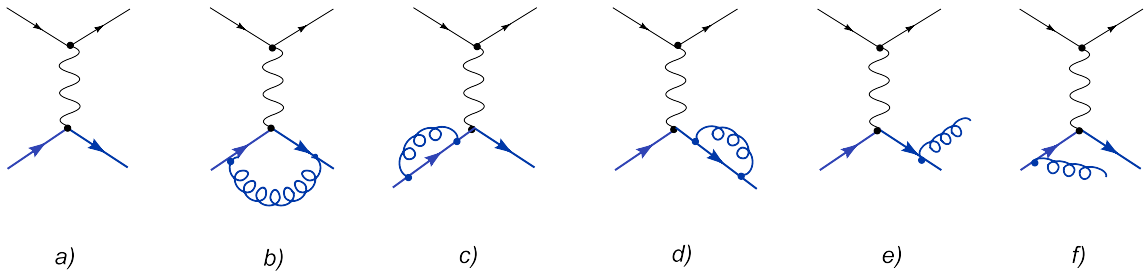}
 \end{center}\vspace{-0.5cm}
 \caption{The process of electron-quark scattering in the first order in  $\alpha_s$: à)
 the Born diagram, b)-d) the corrections due to the virtual gluons,
 e)-f) the corrections due to the real gluons
 }\label{toy}
 \end{figure}

To illustrate the main ideas of the previous chapter, we study first a toy model example.
Let us consider the electron-quark scattering and put all the masses equal to zero. We
will be interested in the radiative corrections in the first order with respect to the
strong coupling $\alpha_s$. The corresponding diagrams are shown in Fig.\ref{toy}.
 We define the initial electron and quark momenta  as  $p_e=p_1$ and  $p_q=p_2$ and  the final ones as $p_{e'}=p_3$ and $p_{q'}=p_4$, respectively.

In the chosen process the UV divergences cancel due to the Ward identities so we are left
only with the IR ones. To handle them, we use dimensional regularization.
 This situation exactly imitates the
four-dimensional CFT's like the  $\mathcal{N}=4$ SYM theory.

Define the measurement function in the following way:
$$
\mathcal{S}_2=\delta_{\pm,h_3}\delta^{D-2} (\Omega_{Det} -
\Omega_{13}),
$$
where $\delta_{\pm,h_3}$ means that we detect the third particle with any helicity, i.e.
we are interested in unpolarized differential cross-section. Here
$d\Omega_{13}=d\phi_{13}d cos(\theta_{13})$\footnote{if to be more accurate in
dimensional regularization we have
$d\Omega_{13}^{D-2}=d\phi_{13}sin(\phi_{13})^{-2\epsilon}d cos(\theta_{13})
sin(\theta_{13})^{-2\epsilon}$, $D=4-2\epsilon$.}, $\theta_{13}$ is the scattering angle
of the particle with momentum $\textbf{p}_3$ with respect to the particle with momentum
$\textbf{p}_1$  in the center of mass frame. In the leading order (LO) we  have the well
known text-book formula \cite{Muta}
\begin{equation}\label{born}
\left(\frac{d\sigma_{2\rightarrow2}}{d\Omega_{13}}\right)_{Born}=
\frac{\alpha^2}{2E^2}\left(\frac{s^2+u^2}{t^2}-\epsilon
\right)\left(\frac{\mu^2}{s}\right)^\epsilon,
\end{equation}
where $E$ is the total energy of initial particles in the center of
mass frame, and $s,t,u$ are the standard Mandelstam variables.
In the c.m. frame  $s=E^2, t=-E^2/2(1-c), u=-E^2/2(1+c)$, $c=\cos \theta_{13}$.

The one loop correction coming from the diagrams with virtual gluon, Fig.\ref{toy} b)-d),
has the form
\begin{equation}\label{virsec}
\left(\frac{d\sigma_{2\rightarrow2}}{d\Omega_{13}}\right)_{1-loop}=
\left(\frac{d\sigma_{2\rightarrow2}}{d\Omega_{13}}\right)_{Born}
\left[-2C_F\frac{\alpha_s}{4\pi}\left(\frac{\mu^2}{-t}\right)^\epsilon (\frac{2}{\mathbf
\epsilon^2}+\frac{3}{\epsilon}+8)\right].
\end{equation}
In order to avoid the transcendental numbers, we used the helpful definition of the
angular measure in the space of $4-2\epsilon$ dimensions and multiplied the standard
expression by $\Gamma(1-\epsilon)/(4\pi)^\epsilon$. Then the constants like $\gamma_E,\
log(4\pi)$ and $\zeta(2)$ disappear from the intermediate expressions. Due to the
cancellation of divergences in the final expressions, this redefinition does not
influence the answer.

Now, following the general prescription, we have to calculate the
diagrams with emission of real gluons, Fig.\ref{toy} e) - f). For
the measurement function of these processes we take
$\mathcal{S}_3=\mathcal{S}_2$, then all requirements on
$\mathcal{S}_2$ and $\mathcal{S}_3$ are satisfied trivially. Besides
the squares of each of the diagrams one should also take into
account the interference term. After contracting all the indices the
phase integral takes the form (we denote the quark momentum as
$p_4-k$ and the gluon momentum as $k$ and keep the standard  notation
for the Mandelstam  variables $s=(p_1+p_2)^2,~$
$t=(p_1-p_3)^2,~$ $u=(p_2-p_3)^2$)
\begin{eqnarray}
\left(\frac{d\sigma_{2\rightarrow3}}{d\Omega_{13}}\right)_{Born}& =&\frac{1}{2\pi
E^2}\int\! d^{D}p_{3} \delta^{+}(p_{3}^{2}) \int\!\! \frac{d^{D}k}{(2\pi)^D}
\delta^{+}(k^{2}) \delta^{+}((p_{4}\!-\!k)^{2}){\cal S}_3
| M |^2_{p_{4}=p_{1}\!+\!p_{2}\!-\!p_{3}},\nonumber\\
&&\hspace*{-2cm} |M|^2=\frac{e^4g^2}{4} 8 \frac{ M_{0} + \epsilon M_{1}+ \epsilon^2 M_{2}
}{t (s+t+u)},\\
&&\hspace*{-2cm} M_{0}= 4 s - 8  p_{1} k - 4 p_{2} k + \frac{- 8 (p_{1} k)^2 +4 (2 s + t)
p_{1} k - (3
s^2  + t^2 + u^2+ 2 s t)}{p_{2} k},\nonumber\\
&& \hspace*{-2cm}M_{1}=  \!-\! 4 (s\!+\!u)\! +\! 8  p_{1} k \!+\! 8 p_{2} k \!+\! \frac{
8 (p_{1} k)^2 \!-\! 4 (s \!+\! t\! +\!u) p_{1}
k\! +\! 2 ( s \!+\! t \!+\! u)^2 \!-\!2(u\!+\!s) t }{p_{2} k}\nonumber, \\
&&\hspace*{-2cm}M_{2}= 4 (s+t+u)  - 4 p_{2} k - \frac{(s+t+u)^2}{ p_{2} k} = -
\frac{(s+t+u -2 p_{2} k)^2}{p_{2} k}.\nonumber
\end{eqnarray}
It is useful to pass to the spherical coordinates and use the c.m. frame. After the
integration over the phase volume the result can be represented as
\begin{eqnarray}
\left(\frac{d\sigma_{2\rightarrow3}}{d\Omega_{13}}\right)_{Born}&=&
 \left(\frac{d\sigma_{2\rightarrow2}}{d\Omega_{13}}\right)_{Born}
\left[2C_F\frac{\alpha_s}{4\pi}\left(\frac{\mu^2}{-t}\right)^\epsilon
(\frac{2}{\epsilon^2}+\frac{3}{\epsilon}+8)\right]\nonumber \\
&+&C_F\frac{\alpha^2}{E^2}\frac{\alpha_s}{4\pi}
\left(\frac{\mu^2}{s}\right)^\epsilon\left(\frac{\mu^2}{-t}\right)^\epsilon
(\frac{f_1}{\epsilon}+f_2)+O(\epsilon),\label{realsec}
\end{eqnarray}
where the functions $f_1$ and $f_2$  in the c.m. frame are
\begin{eqnarray}
f_1&=&-\!2 \frac{(1\!-\!c)(c^3\!+\!5c^2\!-\!3c\!+\!5)\log(\frac{1\!-\!c}{2})\!-\!
(c\!-\!1)^2(c\!+\!1)(c\!-\!11)/4}{(1-c)^2(1+c)^2},\label{f1} \\
f_2&=& -\frac{1}{(1-c)^2(1+c)^2}\left[(1-c)(c^3+5c^2-3c+5)
\log^2(\frac{1-c}{2})\right.\nonumber \\
&&+\left. \frac 12
(1-c)(3c^3\!+\!15c^2\!+\!77c\!-\!31)\log(\frac{1-c}{2})\!+\!(1+c)^2(c^2\!+\!2c\!
+\!5)\pi^2\right. \nonumber
\\&&\left.-12(9c^2\!+\!2c\!+\!5)Li_2(\frac{1+c}{2}) \!+
\!\frac 12(1-c)(1+c)(5c^2\!-\!42c\!-\!23)\right].
\end{eqnarray}

As one can see from comparison of the cross-sections of the processes with virtual
(\ref{virsec}) and real gluons (\ref{realsec}),  in the sum the virtual part {\it
completely} cancels and the second order pole disappears. However, the total cancellation
of divergences does not happen. The remaining divergences in the form of a single pole
have a collinear nature. As was already mentioned, for their cancellation one has to
define properly the initial states.

Introducing the distribution function for initial quark state one
gets the  additional contribution \cite{Alt1} to the cross-section
which looks like\footnote{We put here $\left(\frac{\mu^2}{Q_f^2}\right)^\epsilon$
inside the integration over $z$ since in general one may consider $Q_f$ to be a function of $z$.}
\begin{equation}\label{ColCS}
\left(\frac{d\sigma_{2\rightarrow2}}{d\Omega_{13}}\right)_{Split}=
\frac{1}{\epsilon}\frac{\alpha_s}{2\pi} \int_0^1 dz\ P_{qq}(z)
\left(\frac{\mu^2}{Q_f^2}\right)^\epsilon
\left(\frac{d\sigma_{2\rightarrow2}}{d\Omega_{13}}(p_1,zp_2)\right)_{Born},
\end{equation}
where the Born cross-section is given by (\ref{born}) with the replacement of the initial
quark momentum $p_2$ by $p_2z$.  This means that  one should keep the total
energy but replace the Mandelstam variables  $s,t,u$  according to  eq.(\ref{KinSpl}) (see below).
One should also keep the $\epsilon$ term which gives contributions to the finite part.
The splitting function $P_{qq}(z)$ \cite{Alt} here is
\begin{equation}\label{splitfunc}
P_{qq}(z)=C_F \left(\frac{1+z^2}{(1-z)_+}+\frac 32\delta(1-z)\right).
\end{equation}
Notice also the change of  momenta conservation condition which now
looks like $p_1+zp_2-p_3-p_4=0$. This gives an additional factor of
$4z^{-2\epsilon}/(1+z-c(1-z))^{2(1-\epsilon)}$.

One might also have a contribution  from the final state counterterm; however, since in
this case, according to (\ref{finsp}), the cross-section does not depend on the fraction
$z$, one has to integrate only the splitting function $P_{qq}(z)$. And this integral
equals zero due to the requirement of conservation of the number of quarks. Therefore,
one has no contribution from the final state splitting. It will not be the case for the
gluon scattering cross-section considered below.

The factorization scale $Q_f^2$ is an arbitrary quantity associated with the quark
distribution function which may depend on $z$. It is quite natural to choose the
factorization scale equal to the characteristic scale of the process of interest. Thus,
in our case this choice corresponds to $Q_f^2=-\hat t$, where $\hat t$ is the Mandelstam
parameter $t$ for the process where $p_2$ is replaced by $p_2z$. One has $\hat t =
t\frac{2z}{(z+1)-c(1-z)}$ (see eq.(\ref{KinSpl}) below). Substituting this value of $Q_f^2$ into (\ref{ColCS}) leads to
the following result:
\begin{equation}\label{spl_eq}
\left(\frac{d\sigma_{2\rightarrow2}}{d\Omega_{13}}\right)_{Split}=
C_F\frac{\alpha^2}{2E^2} \frac{\alpha_s}{2\pi}
\left(\frac{\mu^2}{s}\right)^\epsilon
\left(\frac{\mu^2}{-t}\right)^\epsilon (-\frac{f_1}{\epsilon}+f_3),
\end{equation}
where
\begin{eqnarray}
f_3&=& -\frac{1}{(1-c^2)^2}\left[ 2(1-c)(c^3+c^2-33c+7)\log(\frac{1-c}{2})
+12(9c^2+2c+5)Li_2(\frac{1+c}{2})\right.\nonumber
\\&&\left. -(1+c)^2(c^2+2c+5)\pi^2
 -\frac 12(1-c)(1+c)(11c^2-19)\right].
\end{eqnarray}

Gathering all pieces together we finally obtain the IR finite answer   in the NLO order
of PT:
\begin{eqnarray}
  \left(\frac{d\sigma}{d\Omega_{13}}\right)_{IR~safe}&=&\left(\frac{d\sigma_{2\rightarrow2}}{d\Omega_{13}}\right)_{Born}+
  \left(\frac{d\sigma_{2\rightarrow2}}{d\Omega_{13}}\right)_{1-loop}+
  \left(\frac{d\sigma_{2\rightarrow3}}{d\Omega_{13}}\right)_{Born}+
  \left(\frac{d\sigma_{2\rightarrow2}}{d\Omega_{13}}\right)_{Split}\nonumber\\
  && \nonumber \\
&&\hspace*{-2.5cm}=\frac{\alpha^2}{2E^2}\left\{\frac{c^2+2c+5}{(1-c)^2}
-\frac{\alpha_s}{2\pi}\frac{C_F}{(
1-c)(1+c)^2}\left[(c^3+5c^2-3c+5)\log^2(\frac{1-c}{2})\right.\right.
\nonumber\\
&&\hspace*{-1.0cm}\left.\left.+\frac 12 (7c^3+19c^2-55c-3)\log(\frac{1-c}{2})-
(1+c)(3c^2+21c+2)\right]\right\}.
\end{eqnarray}

This expression is the final answer for the cross-section of the physical process of the
electron-quark scattering where the initial and the final state include the soft and
collinear gluons. It includes also the definition of the initial state and can be
recalculated for the alternative choice of the factorization scale similar to what
happens to the ultraviolet scale which defines the coupling constant. Thus, we
practically deal with the scattering not of individual particles but rather with coherent
states with a fixed total momentum. This process contrary to the scattering of individual
massless quanta has a physical meaning. The drawback is the dependence on $Q_f$ which
reflects the definition of the asymptotic state. This dependence explicitly violates the
conformal invariance.

\section{Calculation of the Inclusive Cross-sections in ${\cal N}=4$ SYM theory}\label{s4}

Consider now the gluon scattering in the ${\cal N}=4$ SYM theory. Our aim is to evaluate
the NLO correction to the inclusive differential polarized cross-section in the weak
coupling limit in planar limit in analytical form and to trace the cancellation of the IR
divergences.

We start with the tree level $2 \to 2$ MHV scattering amplitude with
two incoming positively polarized gluons and two outgoing positively
polarized gluons and consider the differential cross-section
$d\sigma_{2 \rightarrow 2}(g^{+}g^{+}\rightarrow
g^{+}g^{+})/d\Omega$ as a function of the scattering solid angle.
Treating all the particles as outgoing this amplitude is denoted as
(-- --  ++) MHV amplitude. At tree level the differential
cross-section is given by
\begin{eqnarray}\label{tree}
  \left(\frac{d\sigma_{2 \rightarrow
2}}{d\Omega_{13}}\right)_{(tree)}^{(--++)} = \frac{1}{J} \int
d\phi_{2} |{\cal M}_{4}^{(tree)}|^2\mathcal{S}_2,
\end{eqnarray}
where $J$ is a flux factor, in our case $J=s$, and the phase volume of the two-particle
process (we use the \textbf{FDH} version of the dimensional reduction, see \cite{FDH} for
details) is
\begin{eqnarray}
 d \phi_{2} =\frac{d^{D} p_{3}\delta^{+}(p_{3}^2)}{(2\pi)^{D-1}} \frac{d^{D} p_{4}
 \delta^{+}(p_{4}^2)}{(2\pi)^{D-1}}
 (2\pi)^D\delta^{D}(p_{1}+p_{2}-p_{3}-p_{4}),
\end{eqnarray}
and $\mathcal{S}_n$ ($n=2$)  in this particular case is
\begin{eqnarray}
\mathcal{S}_2=\delta_{+,h_{3}} \delta^{D-2} (\Omega_{Det} -
\Omega_{13}),
\end{eqnarray}
where $\delta^{D-2} (\Omega_{Det} - \Omega_{13})$ means that our observable is the
differential cross-section and $\delta_{+,h_{3}}$ indicates that we detect a particle
with positive helicity.

The squared matrix element is obtained from the color-ordered amplitudes via summation
\begin{equation}
|{\cal M}^{(tree)}_{4}|^2= g^{4}N_c^{2}(N_c^2-1)\sum_{\sigma \in
P_{3}}|A_4^{(tree)}(p_1,p_{\sigma(2)},p_{\sigma(3)},p_{\sigma(4)})|^2,
\end{equation}
where $P_n$ is the set of the permutations of $n$ objects ($n=3$ in this case), so that
in our  case \cite{Parke,Mangano} (see also appendix A for details)
\begin{equation}\label{4gluonel}
| {\cal M}_{4}^{(tree)(--++)}|^2 \ = \ g^4 N_c^2 (N_c^2-1)
\sum_{\sigma \in P_{3}} \frac{s_{12}^4}{s_{1 \sigma(2)}
s_{\sigma(2)\sigma(3)}s_{\sigma(3) \sigma(4)}s_{\sigma(4) 1}},
\end{equation}
where we use the notation $s_{i j}=(p_{i}+p_{j})^2$.
\begin{figure}[ht]\vspace{0.2cm}
 \begin{center}
 \leavevmode
  \epsfxsize=6cm
 \epsffile{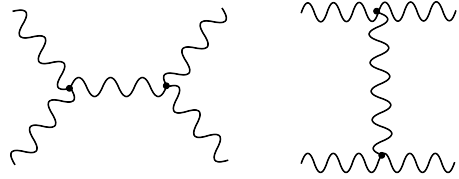}
 \end{center}\vspace{-0.2cm}
 \caption{Tree-level diagrams for the color-ordered MHV amplitudes}\label{glue}
 \end{figure}
 The corresponding Feynman diagrams are shown in Fig.\ref{glue}.

Within the dimensional regularization (reduction) the cross-section in the planar limit
looks like
\begin{equation}
\hspace{-1.5cm}\left(\frac{d\sigma_{2 \rightarrow
2}}{d\Omega_{13}}\right)_{(tree)}^{(--++)}\hspace{-0.3cm} =\frac{\alpha^2N_c^2}{2E^2}
 \left(\frac{s^2}{t^2}+\frac{s^2}{u^2}+\frac{s^4}{t^2u^2}\right)
 \left(\frac{\mu^2}{s}\right)^\epsilon,
 \end{equation}
where $s,t,u$ are the Mandelstam variables, $E$ is the total energy in the center of mass
frame, and  $\alpha = g^2N_c/4\pi$.  So in the center of mass frame the cross-section can
be rewritten as:
\begin{equation}
\hspace{-1.5cm}\left(\frac{d\sigma_{2 \rightarrow
2}}{d\Omega_{13}}\right)_{(tree)}^{(--++)}\hspace{-0.3cm}
=\frac{\alpha^2N_c^2}{E^2}\frac{4(3+c^2)}{(1-c^2)^2}\left(\frac{\mu^2}{s}\right)^\epsilon,
\end{equation}
where $c=\cos \theta_{13}$. The next step is to calculate the NLO corrections.

\subsection{Virtual part}
\begin{figure}[ht]\vspace{0.2cm}
 \begin{center}
 \leavevmode
  \epsfxsize=15cm
 \epsffile{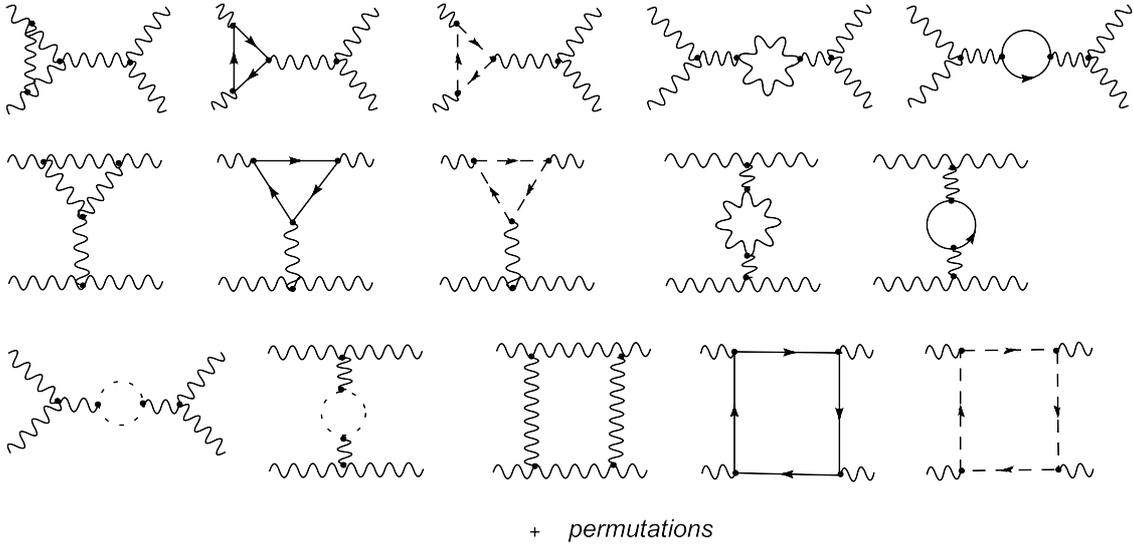}
 \end{center}\vspace{-0.5cm}
 \caption{The one-loop diagrams for the color-ordered MHV amplitude in the ${\cal N}=4$ SYM theory.
Particles running inside the loop include all the members of the ${\cal N}=4$
supermultiplet. The solid and dashed lines correspond to the fermion and scalar
particles, respectively. }\label{gluons}
 \end{figure}
To get the one-loop contribution to the differential cross-section,
one has to consider the diagrams shown in Fig.\ref{gluons}. We use
the already known one loop  contribution to the color-ordered
amplitude~\cite{Green}
$$M_4^{(1-loop)}(\epsilon)=A_4^{(1-loop)}/A_4^{(tree)}\ = - \frac 12 st I_4^{(1-loop)}(s,t),$$
where $I_4^{(1-loop)}(s,t)$ is the scalar box diagram
$$ I_4^{(1-loop)}(s,t) \ = \
\frac{2}{st}\frac{\Gamma(1\!-\!\epsilon)^2}{\Gamma(1-2\epsilon)}
[-\frac{1}{\epsilon^2}\left((\frac{\mu^2}{s})^{\epsilon}\!\!+\!(\frac{\mu^2}{-t})^{\epsilon}\right)
\!+\!\frac{1}{2}\log^2\left(\frac{s}{-t}\right)\!+\!\frac{\pi^2}{3}]+\mathcal{O}(\epsilon).
$$
The square of the matrix element summed over colors
$$|{\cal
M}_{4}^{(1-loop)}|^2=\sum_{colors}(\mathcal{A}_4^{(tree)}
\mathcal{A}_4^{(1-loop)\ast}+c.c.)$$ has the form
\begin{eqnarray}\label{4gluonel1l}
|{\cal M}_{4}^{(1-loop)(--++)}|^2 &=& - g^4 N_c^2 (N_c^2-1)
\left(\frac{g^2N_c}{16\pi^2}\right) \\
&&\hspace{-3cm}\times \left[ \frac{s^4}{s^2t^2} st I_4^{(1-loop)}(s,t) +
\frac{s^4}{s^2u^2} su I_4^{(1-loop)}(s,u)  - \frac{s^4}{t^2u^2} tu I_4^{(1-loop)}(-t,u)
\right],\nonumber
\end{eqnarray}
which gives the one-loop contribution to the cross-section in the planar limit
\begin{eqnarray}
 && \hspace{-1.5cm}\left(\frac{d\sigma_{2 \rightarrow 2}}{d\Omega_{13}}\right)_{virt}^{(--++)}
 =\frac{\alpha^2N_c^2}{2E^2}
 \left(\frac{\mu^2}{s}\right)^\epsilon\left\{\frac{\alpha}{4\pi}
 \frac{s^4}{s^2 t^2 u^2}\left[- \frac{8}{\epsilon^2}\left(
((\frac{\mu^2}{-t})^\epsilon+(\frac{\mu^2}{-u})^\epsilon)s^2\right.\right. \right. \nonumber \\
&&\hspace{-1.2cm}\left.\left.\left. +
((\frac{\mu^2}{s})^\epsilon+(\frac{\mu^2}{-t})^\epsilon)u^2+
((\frac{\mu^2}{s})^\epsilon+(\frac{\mu^2}{-u})^\epsilon)t^2\right)\right.\right.  \nonumber \\
&& \hspace{-1.2cm}\left.\left. +
\frac{16}{3}\pi^2(s^2\!+\!t^2\!+\!u^2)+4(u^2\log^2(\frac{s}{-t}) +
t^2\log^2(\frac{s}{-u}) + s^2\log^2(\frac{t}{u}))\right]\right\}.
\end{eqnarray}
Rewriting this expression in the center of mass frame we have:
\begin{eqnarray}\label{1loop4gluon}
&& \hspace*{-1cm}\left(\frac{d\sigma_{2 \rightarrow
2}}{d\Omega_{13}}\right)_{virt}^{(--++)}=\frac{\alpha^2 N_c^2}{E^2}
 \left(\frac{\mu^2}{s}\right)^{2\epsilon}4\left\{\frac{\alpha }{4\pi}
 \left[-\frac{16}{\epsilon^2}\frac{3+c^2}{(1-c^2)^2}
+ \frac{4}{\epsilon}
\left(\frac{5+2c+c^2}{(1-c^2)^2}\log(\frac{1-c}{2})\right.\right.\right. \nonumber \\
&&\hspace{-1cm}\left.\left.\left.+\frac{5-2c+c^2}{(1-c^2)^2}\log(\frac{1+c}{2})
\phantom{\frac{1+c}{2}}\hspace{-0.7cm}\right) +\frac{16(3+c^2)\pi^2}{3(1-c^2)^2}-
\frac{16}{(1-c^2)^2}\log(\frac{1-c}{2})\log(\frac{1+c}{2})\right]\right\}.
\end{eqnarray}

It should be  stressed that due to the conformal invariance of the $\mathcal{N}=4$ SYM
theory at the quantum level there are no UV divergences in (\ref{1loop4gluon}) and all
divergences have the IR soft or collinear nature. They have to be canceled in properly
defined observables. Note also  the simplicity of the finite part which is a consequence
of symmetries of $\mathcal{N}=4$ SYM and the fact  that all the terms in
(\ref{1loop4gluon}) have the same transcendentality \cite{BES06,KotLip}.

\subsection{Real emission}

\begin{figure}[ht]\vspace{0.2cm}
 \begin{center}
 \leavevmode
  \epsfxsize=14cm
 \epsffile{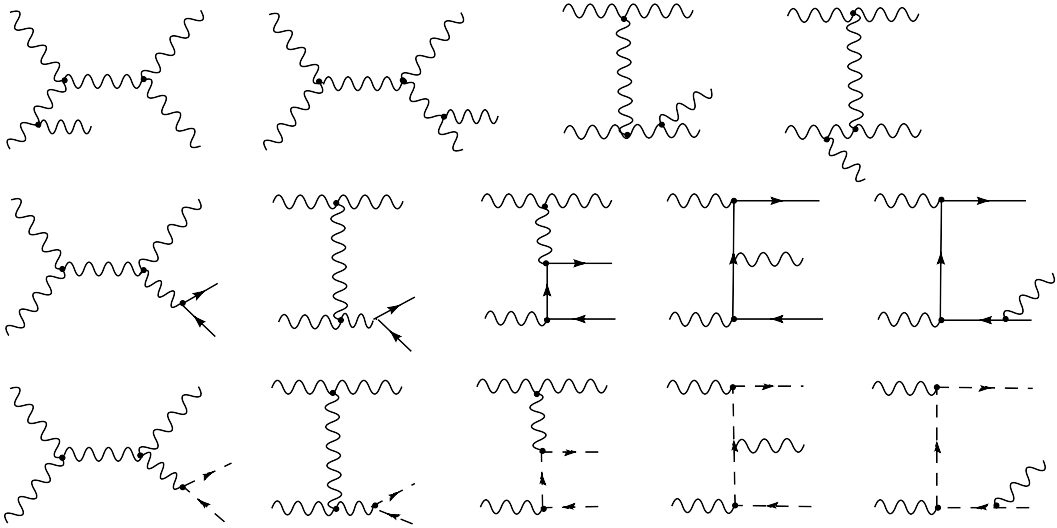}
 \end{center}\vspace{-0.2cm}
 \caption{The tree diagrams with three outgoing particles for the color-ordered
 amplitudes. Permutations are not shown.
 \label{real}}
 \end{figure}

The next step, as in the toy model considered above, is the calculation of the amplitude
with three outgoing particles. Here we have to define  the process  we are interested in.
There are several possibilities.
\begin{enumerate}
  \item Three gluons with positive helicities: $g^+g^+ \rightarrow g^+g^+g^+.$
  This is the MHV amplitude;
  \item Two gluons with  positive helicities and the
  third one with negative helicity: $g^+g^+ \rightarrow
  g^+g^+g^-.$\footnote{There is also a $g^+g^+ \rightarrow
> g^+g^-g^+$ helicity configuration. The partial amplitudes for both
> cases where the additional gluon with negative helicity is the
> second or the third gluon in the final state are equal. We will use
> the $(--++-)$ notation for both of them.}
  This is the anti-MHV amplitude;
  \item One of three final particles is the gluon with positive
  helicity and the rest is the quark-antiquark pair\footnote{The $\mathcal{N}=4$
  supermultiplet
  consists of a gluon $g$, 4 fermions ("quarks") $q^{A}$ and 6 real scalars
  $\Lambda^{AB}$; $A$ and $B$ are $SU(4)_{R}$ indices, $\Lambda$ is an antisymmetric
  tensor. It is implied that all squared  amplitudes with quarks
  and scalars are summed over these indices.}:
  $g^+g^+ \rightarrow \ g^+q^-\overline{q}^+$ or $g^+g^+ \rightarrow g^+q^+\overline{q}^-.$
  This  is an anti-MHV amplitude;
  \item One of three final particles is the gluon with positive
  helicity and the rest are  two scalars:
  $g^+g^+ \rightarrow g^+ \Lambda \Lambda.$ This  is an
  anti-MHV amplitude.
\end{enumerate}
The corresponding diagrams are shown in Fig.\ref{real}.

If one fixes one gluon with positive helicity scattered at angle
$\theta$ and sums over all the other particles then all the
processes mentioned above contribute.  In the case when one fixes
two gluons with positive helicity and looks for the rest, only the
first two options are allowed.

The cross-section of these processes can be written as
\begin{eqnarray}\label{crossf}
 \left(\frac{d\sigma_{2 \rightarrow
3}}{d\Omega_{13}}\right)_{Real} =  \frac{1}{J} \int d\phi_{3} |
{\cal
 M}_5^{(tree)}|^2 \mathcal{S}_3,
\end{eqnarray}
where $d\phi_{3}$ is the three-particle phase volume
\begin{eqnarray}
 d \phi_{3} =\frac{d^{D} p_{3}\delta^{+}(p_{3}^2)}{(2\pi)^{D-1}}
 \frac{d^{D} p_{4}\delta^{+}(p_{4}^2)}{(2\pi)^{D-1}}
 \frac{d^{D} p_5 \delta^{+}(p_5^2)}{(2\pi)^{D-1}}
 (2\pi)^D\delta^{D}(p_{1}+p_{2}-p_{3}-p_{4}-p_{5}),
\end{eqnarray}
and $\mathcal{S}_3$ is the measurement function which constraints the phase space and
defines the particular observable.

The squared matrix element is expressed through the amplitudes as before
\begin{equation}
|{\cal M}^{(tree)}_5(p_1,...,p_5)|^2 =
g^{6}N_c^{3}(N_c^2-1)\sum_{\sigma \in
P_{4}}|A_5^{(tree)}(p_1,p_{\sigma(1)},...,p_{\sigma(4)})|^2.
\end{equation}

For the processes mentioned above one has the following expressions for the matrix
elements: \footnote{It is implied that all squared  amplitudes with quarks
  and scalars are summed over $SU(4)_{R}$ indices.}:
  \begin{eqnarray}\label{5gl}\hspace*{-1cm}1.\hspace{1.0cm} |{\cal M}_5^{(tree)(--+++)}|^2=g^6 N_c^3 (N_c^2-1) \sum_{\sigma
\in P_{4}} \frac{s_{12}^4}{s_{1 \sigma(1)}
s_{\sigma(1)\sigma(2)}s_{\sigma(2)\sigma(3)}s_{\sigma(3)
\sigma(4)}s_{\sigma(4) 1}}.
\end{eqnarray}
Since there are three identical gluons with positive helicity in the final state  one has
to define which ones are detected. In case of one detectable particle, one can choose the
fastest one; in case of two, the two fastest ones. The measurement function  for
detecting only one gluon with momentum $p_3$ with positive helicity can be written as
\begin{eqnarray}\label{mf3p1a}
{\cal S}^{(--+++),1}_{3} = 
\delta_{+,h_{3}} \Theta(p^0_{3} > p^0_{4}) \Theta(p^0_{3} >
p^0_{5})\delta^{D-2} (\Omega_{Det} - \Omega_{13});
\end{eqnarray} and for detecting of two gluons with positive
helicities as
\begin{eqnarray}\label{mf3p1}
{\cal S}^{(--+++),2}_{3} =  \delta_{+,h_{3}} \delta_{+,h_{4}}\Theta(p^0_{3} > p^0_{5})
\Theta(p^0_{4} > p^0_{5})\delta^{D-2} (\Omega_{Det} - \Omega_{13}) ,\end{eqnarray} where
we detect the 3-rd and the 4-th gluons. Analogous measurement function would appear if we
would like to detect the 3-rd and the 5-th gluons.

\begin{eqnarray}\label{4gl}\hspace*{-1cm}2.\hspace{1.0cm}
 |{\cal M}_5^{(tree)(--++-)}|^2=g^6 N_c^3 (N_c^2-1) \sum_{\sigma \in P_{4}} \frac{s_{34}^4}{s_{1 \sigma(1)} s_{\sigma(1)\sigma(2)}s_{\sigma(2)\sigma(3)}s_{\sigma(3) \sigma(4)}s_{\sigma(4)
 1}};
\end{eqnarray}
The measurement function for detecting one gluon with positive helicity and momentum
$p_3$ is given by
\begin{eqnarray}\label{mf3p1b}
{\cal S}^{(--+++),1}_{3} = 
\delta_{+,h_{3}} \Theta(p^0_{3} > p^0_{4}) )\delta^{D-2} (\Omega_{Det} - \Omega_{13});
\end{eqnarray}
and for detecting of two gluons with positive helicity  by
\begin{eqnarray}\label{mf3p2}
{\cal S}^{(--++-),2}_{3} = 
\delta_{+,h_{3}} \delta_{+,h_{4(5)}} \delta^{D-2} (\Omega_{Det} -
\Omega_{13}).\end{eqnarray}
 \begin{eqnarray}\label{qq}\hspace*{-1cm}3.\hspace{1.0cm}
|{\cal M}_5^{(tree)(--+q\overline{q})}|^2=g^6 N_c^3 (N_c^2-1) \sum_{\sigma \in P_{4}}
\frac{s_{34}s_{35} (s^2_{34}+s^2_{35}) }{s_{1 \sigma(1)}
s_{\sigma(1)\sigma(2)}s_{\sigma(2)\sigma(3)}s_{\sigma(3) \sigma(4)}s_{\sigma(4)1}};
\end{eqnarray}
The measurement function  in this case is simple since we have only one gluon in the
final state
\begin{eqnarray}\label{mf3p3}
{\cal S}^{(--+q\overline{q})}_{3} = \delta_{+,h_{3}}  \delta^{D-2} (\Omega_{Det} -
\Omega_{13}) .\end{eqnarray}
 \begin{eqnarray}\label{scalar}\hspace*{-1cm}4.\hspace{1.0cm}
  |{\cal M}_5^{(tree)(--+\Lambda\Lambda)}|^2= g^6 N_c^3 (N_c^2-1) \sum_{\sigma \in P_{4}}
  \frac{s_{34}^2 s_{35}^2}{s_{1 \sigma(1)} s_{\sigma(1)\sigma(2)}s_{\sigma(2)\sigma(3)}s_{\sigma(3) \sigma(4)}s_{\sigma(4)
  1}}.
\end{eqnarray}
The measurement function is given by the same formula (\ref{mf3p3}) as in the previous
case.

One can check that the measurement functions written above  satisfy the IR and collinear
limit conditions (\ref{appropr2},\ref{appropr}). Indeed, one has

 1. $p_5\to 0,|\mathbf{p}_4|=|\mathbf{p}_3|$:
$$S_3(p_3,p_4,0)\to \Theta(p_3^0-p_4^0)\delta^{D-2}(\Omega-\Omega_3),$$

2. $\mathbf{p}_3=-\mathbf{P},\mathbf{p}_4=x\mathbf{P},\mathbf{p}_5=(1-x)\mathbf{P}$:
 $$ S_3(p_3,p_4,p_5)\to
\Theta(1-x)\Theta(x)\delta^{D-2}(\Omega-\Omega_3).$$ The latter $\theta$-functions give
$0<x<1$ restricting the fraction of momenta in a natural way.

Choosing the fastest momentum one has to have in mind the conservation of momentum and
energy
$$ \mathbf{p}_3+\mathbf{p}_4 +\mathbf{p}_5=0,\ \ \ |\mathbf{p}_3|+|\mathbf{p}_4|+
|\mathbf{p}_5|=E.$$ This means that the three momenta form a
triangle with the perimeter equal to E. Hence the requirement that,
say, the third particle is the fastest one means that $p_3^0>E/3$.
Therefore, to simplify the integration, in what follows we choose
the universal measurement function
\begin{equation}\label{mf} \mathcal{S}_3(p_3,p_4,p_5) =
 \Theta(p_3^0 - \frac{1 - \delta}{2} E)
\delta^{D-2}(\Omega_{Det}-\Omega_{3}),
\end{equation}
where we take $\delta =1/3$ in the case of identical particles and   $\delta=1$ in the
other cases. Thus, the registration of one fastest gluon corresponds to $\delta =1/3$ for
the MHV and anti-MHV amplitudes  and $\delta=1$ for the matter-antimatter amplitude,
while the registration of two fastest gluons corresponds to  $\delta =1/3$ for the MHV
amplitude and  $\delta =1$ for the anti-MHV amplitude\footnote{These are not precisely the needed requirements  but are pretty close to them. Fulfillment of the exact requirements of the fastest particles is technically more involved but does not changed the picture.}.
In what follows we  keep the value of $\delta$ arbitrarily and show that the IR and
collinear divergences cancel in observables for any value of $\delta$. We omit the
details of the calculation, which can be found in Appendix B, and present here only the
divergent parts of the calculated objects. All the finite parts can be found in Appendix
D.

With these definitions the contributions to the $2\to 3$ cross-sections from the
amplitudes that are listed above  are

1. Real Emission (MHV)
\begin{eqnarray}
&&\hspace{-0.5cm}\left(\frac{d\sigma_{2 \rightarrow
3}}{d\Omega_{13}}\right)^{(--+++)}_{Real} =\frac{\alpha^2N_c^2}{E^2}
 \left(\frac{\mu^2}{s}\right)^{2\epsilon}\frac{\alpha}{\pi}\left\{
\frac{8}{\epsilon^2}\frac{(3+c^2)}{(1-c^2)^2} \right.  \\
&&\hspace{-0.5cm}\left. + \frac{1}{\epsilon}\left[
\frac{2}{(1\!+\!c)^2}\log(\frac{1\!-\!c}{2})\!+\!\frac{2}{(1\!-\!c)^2}\log(\frac{1\!+\!c}{2})+
\frac{16\delta(2\delta\!-\!3)}{(1\!-\!c^2)^2 (1\!-\!\delta)^2} \!+\!
\frac{12(3\!+\!c^2)}{(1\!-\!c^2)^2}\log(\frac{1\!-\!\delta}{\delta})\right] \right.
\nonumber
\\&& \hspace{-0.5cm}\left.
+\mbox{Finite part}\right\};\nonumber
\end{eqnarray}
Notice the singularity in the limit $\delta\to 1$.

2. Real Emission (anti-MHV)
\begin{eqnarray}
&&\hspace{-0.0cm}\left(\frac{d\sigma_{2 \rightarrow
3}}{d\Omega_{13}}\right)^{(--++-)}_{Real} =\frac{\alpha^2 N_c^2}{E^2}
 \left(\frac{\mu^2}{s}\right)^{2\epsilon}\frac{\alpha}{\pi}\left\{
\frac{1}{\epsilon^2}\frac{8(3+c^2)}{(1-c^2)^2}+\frac{1}{\epsilon}\left[
-\frac{12(c^2+3)\log\delta}{(1-c^2)^2}\right.\right. \nonumber\\
&&\hspace{1.3cm}+\left. \left.  \frac{64 (12 c^2+17)}{3(1-c^2)^3} +
\frac{2\delta}{(1-c^2)^2} \left( \frac 23
(5+3c^2) \delta^2 - (c^2+19) \delta + 2 (5c^2+43) \right) \right. \right.\nonumber\\
&&\hspace{1.3cm}+\left.\left.
\left(\frac{2(3c^2-24c+85)}{(1-c)(1+c)^3}\log(\frac{1-c}{2})-
\frac{8(c^2-6c+21)}{(1-c)(1+c)^3}\log(\frac{1+\delta\!-\!(1-\delta)c}{2})
\right. \right. \right. \nonumber\\
&& \hspace{1.3cm}\left.
\left.\left.-\frac{32(c^2-4c+7)}{(1+c)^3(1-c)(1+\delta-c(1-\delta))}+
\frac{32(2-c)}{(1+c)^3(1+\delta-c(1-\delta))^2}\right. \right. \right.\nonumber\\
&&\hspace{1.3cm}\left. \left.\left. -\frac{64(1-c)}{3(1+c)^3(1+\delta-c(1-\delta))^3}+
(c\leftrightarrow -c) \right)\right]+ \mbox{Finite part} \right\};
\end{eqnarray}

Contrary to the MHV case  the limit $\delta \to 1$ is regular here and greatly simplifies
the final result.

3. Fermions (for 4 fermions in adjoint representation of $SU(N_c)$)
\begin{eqnarray}
&&\left(\frac{d\sigma_{2 \rightarrow 3}}{d\Omega_{13}}\right)^{(--+q{\bar q})}_{Real}
=\frac{\alpha^2 N_c^2}{E^2}
 \left(\frac{\mu^2}{s}\right)^{2\epsilon}\frac{\alpha}{\pi}\left\{-
\frac{16}{\epsilon}\left[\frac{(79+25c^2)}{3(1-c^2)^2}\right.\right.
\\ \nonumber &&\hspace{-0.5cm}\left.\left.+
\frac{2(3-c)^2}{(1-c)(1+c)^3}\log(\frac{1-c}{2})
+\frac{2(3+c)^2}{(1-c)^3(1+c)}\log(\frac{1+c}{2})\right]+\mbox{Finite part}\right\};
\end{eqnarray}

 4. Scalars (for 6 scalars in adjoint representation of $SU(N_c)$)
\begin{eqnarray}
&&\hspace{-0.5cm} \left(\frac{d\sigma_{2 \rightarrow
3}}{d\Omega_{13}}\right)^{(--+\Lambda\Lambda)}_{Real} =\frac{\alpha^2 N_c^2}{E^2}
 \left(\frac{\mu^2}{s}\right)^{2\epsilon}\frac{\alpha}{\pi}\left\{-
\frac{8}{\epsilon}\left[-\frac{2(10+7c^2)}{(1-c^2)^2}\right.\right.
\\  \nonumber &&\hspace{-0.5cm}\left.\left.-
\frac{3(5-c)}{(1+c)^3}\log(\frac{1-c}{2})-\frac{3(5+c)}{(1-c)^3}
\log(\frac{1+c}{2})\right] +\mbox{Finite part}\right\}.
\end{eqnarray}

In the last two expressions we chose the parameter $\delta=1$ since there are no
identical particles in these cases and there is no need to restrict the phase space. Note
also the absence of the second order pole in $\epsilon$ which means that there is no IR
soft divergency here but  only a collinear one.

\subsection{Splitting}

\begin{figure}[ht]\vspace{0.3cm}
 \begin{center}
 \leavevmode
  \epsfxsize=15cm
 \epsffile{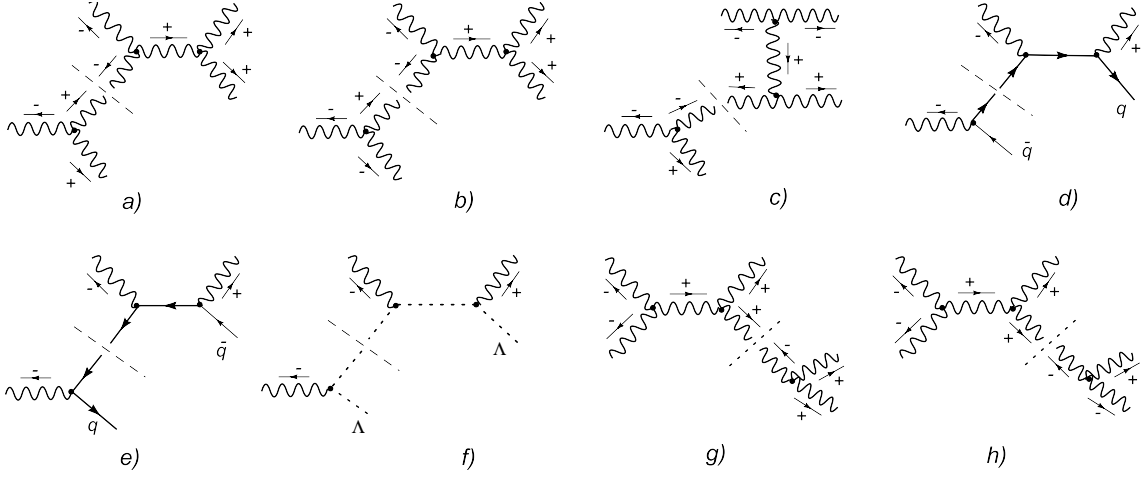}
 \end{center}\vspace{-0.2cm}
 \caption{The initial and final particle splitting diagrams:
 a) the initial MHV amplitude, b)-c) the initial anti-MHV amplitudes, d)-f) the initial
 matter amplitudes, g)-h) the final MHV and anti-MHV amplitudes.
 Permutations are not shown. \label{split}}
 \end{figure}

Now we have to deal with an additional $1/\epsilon$ pole coming from the collinear
divergences. As one can see from the  toy model example, taking into account emission of
additional quanta in the initial and  final states allows one to cancel the IR
divergences (double poles in $\epsilon$) but leaves the single poles originating from
collinear ones. Indeed,  as it has been discussed earlier, the asymptotic states (both
the initial and final ones) are not well defined since a massless quantum can split into
two parallel ones indistinguishable from the original. To take this into account, we
introduce the distribution of the initial and final particle (gluon or any other member
of the $\mathcal{N}=4$ SYM supermultiplet) with respect to the fraction of the carried
momentum $z$: $q_i(z,Q_f^2 / \mu^2)$. Also, one has to keep in mind that the particles in
this case are polarized. The corresponding Feynman diagrams are shown in Fig.\ref{split}.

Additional contributions from collinear particles in the initial or final states to the
inclusive gluon cross-section (the collinear counterterms) have the following form,
respectively:
\begin{eqnarray}\label{AP}
d \sigma_{2 \rightarrow 2}^{spl,init} = \frac{\alpha}{2\pi}\frac{1}{\epsilon}
\left(\frac{\mu^2}{Q^2_f}\right)^\epsilon \sum\limits_{i,j=1,2;\ i\neq j}\int_{0}^{1} d z
\!\!\!\sum\limits_{l=g,q,\Lambda}\!\!P_{gl}(z)d \sigma_{2 \rightarrow 2} (z
p_{i},p_{j},p_{3},p_{4}){\cal S}_2^{spl,init}(z),
\end{eqnarray}
\begin{eqnarray}\label{APfin}
d \sigma_{2 \rightarrow 2}^{spl,fin} = \frac{\alpha}{2\pi}\frac{1}{\epsilon}
\left(\frac{\mu^2}{Q^2_f}\right)^\epsilon d \sigma_{2 \rightarrow 2}
(p_{1},p_{2},p_{3},p_{4}) \int_{0}^{1} d z\sum\limits_{l=g,q,\Lambda}P_{gl}(z) {\cal
S}_2^{spl,fin}(z).
\end{eqnarray}

Having particles with different helicities  we have the following set of collinear
counterterms (We use here slightly different notation for the splitting functions
indicating explicitly all three particles like $P^{init}_{fin_1,fin_2}(z)$ to avoid
confusion.)

1. Initial state splitting MHV amplitude $(--+++)$
\begin{eqnarray}\label{Spl3gl+}
&&\hspace*{-1.5cm} \left(\frac{d\sigma_{2 \rightarrow
2}}{d\Omega_{13}}\right)^{(--+++)}_{InSplit} = \frac{\alpha}{2\pi}\frac{1}{\epsilon}
\left(\frac{\mu^2}{Q^2_f}\right)^\epsilon \int_{0}^{1}\! d z\ 2\ P_{g^+g^+}^{g^-}(z)
\left(\frac{d \sigma_{2 \rightarrow
2}}{d\Omega_{13}}\right)^{(--++)}\hspace{-1.2cm}(zp_1,p_2,p_3,p_4)  {\cal S}_{2,\
init}^{(--+++)}(z)\nonumber \\&& \hspace*{1.8cm}+\  (p_1 \leftrightarrow p_2),
\end{eqnarray}

Final state splitting MHV amplitude $(--+++)$
\begin{equation}
 \left(\frac{d\sigma_{2 \rightarrow
2}}{d\Omega_{13}}\right)^{(--+++)}_{FnSplit} = 2\frac{\alpha}{2\pi}\frac{1}{\epsilon}
\left(\frac{\mu^2}{Q^2_f}\right)^\epsilon \left(\frac{d \sigma_{2 \rightarrow
2}}{d\Omega_{13}}\right)^{(--++)}\hspace{-1.2cm}(p_1,p_2,p_3,p_4) \int_{0}^{1}\! d z\
P_{g^+g^+}^{g^-}(z) {\cal S}_{2,\ fin}^{(--+++)}(z).
\end{equation}

2. Initial state splitting anti-MHV amplitude $(--++-)$
\begin{eqnarray}\label{Spl3gl-}
 \left(\frac{d\sigma_{2 \rightarrow
2}}{d\Omega_{13}}\right)^{(--++-)}_{InSplit} &=& \frac{\alpha}{2\pi}\frac{1}{\epsilon}
\left(\frac{\mu^2}{Q^2_f}\right)^\epsilon \int_{0}^{1} dz\ 2\ P_{g^-g^+}^{g^-}(z)
\left(\frac{d \sigma_{2 \rightarrow
2}}{d\Omega_{13}}\right)^{(-++-)}\hspace{-1.2cm}(zp_1,p_2,p_3,p_4) {\cal S}_{2,\
init}^{(--++-)}(z)  \nonumber \\&+&\frac{\alpha}{2\pi}\frac{1}{\epsilon}
\left(\frac{\mu^2}{Q^2_f}\right)^\epsilon \int_{0}^{1}\! d z\ 2\ P_{g^+g^-}^{g^-}(z)
\left(\frac{d \sigma_{2 \rightarrow
2}}{d\Omega_{13}}\right)^{(--++)}\hspace{-1.2cm}(zp_1,p_2,p_3,p_4) {\cal
S}_{2,\ init}^{(--++-)}(z)  \nonumber \\
 &+&(p_1 \leftrightarrow p_2),
\end{eqnarray}

Final state splitting anti-MHV amplitude $(--++-)$
\begin{equation}
 \left(\frac{d\sigma_{2 \rightarrow
2}}{d\Omega_{13}}\right)^{(--++-)}_{FnSplit} = 2\frac{\alpha}{2\pi}\frac{1}{\epsilon}
\left(\frac{\mu^2}{Q^2_f}\right)^\epsilon \left(\frac{d \sigma_{2 \rightarrow
2}}{d\Omega_{13}}\right)^{(--++)}\hspace{-1.2cm}(p_1,p_2,p_3,p_4)
 \int_{0}^{1}\! d z\ P_{g^+g^-}^{g^-}(z) {\cal
S}_{2,\ fin}^{(--++-)}(z).
\end{equation}

One has also the collinear counterterms containing the other members of the
$\mathcal{N}=4$ supermultiplet

3. Initial state splitting into a fermion-antifermion pair
\begin{eqnarray}\label{SplQuark}
&&\hspace*{-0.7cm} \left(\frac{d\sigma_{2 \rightarrow
2}}{d\Omega_{13}}\right)^{(--+q\overline{q})}_{InSplit} =
\frac{\alpha}{2\pi}\frac{n_f}{\epsilon}
\left(\frac{\mu^2}{Q^2_f}\right)^\epsilon \int_{0}^{1}\! d z
\left[2\ P_{\bar{q}^-q^+}^{g^-}(z) \left(\frac{d \sigma_{2
\rightarrow 2}}{d\Omega_{13}}\right)^{(q-+\bar
q)}\hspace{-0.8cm}(zp_1,p_2,p_3,p_4)  {\cal S}_{2,\
init}^{(--+q\overline{q})}(z) \right. \nonumber \\ && \hspace{2.4cm}
\left.
 + 2\ P_{q^+\bar{q}^-}^{g^-}(z) \left(\frac{d \sigma_{2 \rightarrow
2}}{d\Omega_{13}}\right)^{{(\overline{q}-+q)}}
\hspace{-0.8cm}(zp_1,p_2,p_3,p_4)
  {\cal S}_{2,\ init}^{(--+q\overline{q})}(z)
 \right]+ (p_1 \leftrightarrow p_2),
\end{eqnarray}

4. Initial state splitting into  a scalar pair
\begin{eqnarray}\label{SplScalar}
&&\hspace*{-0.8cm}\left(\frac{d\sigma_{2 \rightarrow
2}}{d\Omega_{13}}\right)^{(--+\Lambda\Lambda)}_{InSplit} =
\frac{\alpha}{2\pi}\frac{n_s}{\epsilon}
\left(\frac{\mu^2}{Q^2_f}\right)^\epsilon \int_{0}^{1} dz\ 2\
P_{\Lambda\Lambda}^{g^-}(z)\left(\frac{d \sigma_{2 \rightarrow
2}}{d\Omega_{13}}\right)^{{(\Lambda- +\Lambda)}}
\hspace{-1.1cm}(zp_1,p_2,p_3,p_4)
 {\cal S}_{2,\ init}^{(--+\Lambda\Lambda)}(z)\nonumber \\&&\hspace*{2.4cm} + \ (p_1
\leftrightarrow p_2),
\end{eqnarray}
where $n_f$ and $n_s$ is the number of fermions and scalars, respectively. One should put
$n_f=4$ and $n_s=6$ in our case.

The explicit form of the Born cross-sections and the splitting
functions $P^i_{jk}(z)$ can be found in Appendices A and C,
respectively. Notice that when changing the momentum $p_i \to zp_i$
one has to modify the Mandelstam variables according to
eq.(\ref{KinSpl}) and take into account the additional factor from
the phase space in full analogy with the QED case (see the comment
after eq.(\ref{ColCS})).

Note that there is no final state splitting counterterms for fermions and scalars. The
reason is that one has to take into account only those final splittings where the
original state (gluon in our case) survives with momentum multiplied by fraction $z$.

The measurement functions here  are the same as in the case of real
emission but depend now on fraction $z$ and restrict the integration
region over $z$. They  take the form
\begin{equation}\label{sz}
{\cal S}_2^{spl,1}(z) = \delta_{+,h_3}
  \delta^{D-2} (\Omega-\Omega_{13})
  \Theta(z-z_{min}),
\end{equation} for detecting of one gluon and
\begin{equation}
  {\cal S}_2^{spl,2}(z) = \delta_{+,h_{3}} \delta_{+,h_{4,5}}
  \delta^{D-2} (\Omega-\Omega_{13})\Theta(z-z_{min})
  \end{equation}
for detecting of two gluons.

The values of $z_{min}$ can be calculated from the requirement $p_3^0>(1-\delta)E/2$ in
different kinematics. Indeed, for the initial splitting process one has to change the
momentum of in-going particle, for example,  $p_1$ to  $z p_{1}$ which gives in the c.m.
frame
\begin{eqnarray}\label{KinSpl}
&s\to z s, \nonumber \\
&t\to\frac{2 z}{1+ z - c (1 - z)} t, \nonumber \\
&u\to\frac{2 z^2}{1+ z - c (1 - z)} u, \nonumber \\
&p_{3}^{0}\to \frac{2 z}{1+ z - c (1 - z)} \frac{E}{2}.
\end{eqnarray}
At the same time, for the final splitting one has to substitute $p_3^0\to z\frac E2$.
This leads to the  values of $z_{min}$, respectively,
\begin{equation}\label{Up1}
z_{min}^{in} \ = \ \frac{(1 - \delta) (1 - c)}{1 + \delta - c (1 -
\delta)}, \ \ \ z_{min}^{fin} \ = \ (1 - \delta).
\end{equation}

 Taking into account the splitting of the initial states and the fragmentation of the
final states we get the following contribution to the inclusive cross-sections:
\begin{enumerate}
  \item
The initial  and final splitting for the MHV amplitude
\begin{eqnarray}
&&\hspace{-1cm}\left(\frac{d\sigma_{2 \rightarrow
3}}{d\Omega_{13}}\right)^{(--+++)}_{InSplit} =\frac{\alpha^2 N_c^2}{E^2}
 \left(\frac{\mu^2}{s}\right)^{\epsilon}\left(\frac{\mu^2}{Q_f^2}\right)^{\epsilon}\!\!
 \frac{\alpha}{\pi}\left\{
\frac{1}{\epsilon}\left[
-\frac{4(c^2\!+\!3)}{(1\!-\!c^2)^2}\left(\log\frac{1\!-\!c}{2}+\log\frac{1\!+\!c}{2}\right)
\right. \right. \nonumber \\ && \left. \left. \makebox[1em]{} -\frac{8(c^2+3)}{(1-c^2)^2}
\log\frac{1-\delta}{\delta} -\frac{16\delta(2\delta-3)}{(1-c^2)^2(1-\delta)^2} \right]+
\mbox{Finite part}\right\},
\end{eqnarray}
\begin{eqnarray}
&& \hspace{-3cm}\left(\frac{d\sigma_{2 \rightarrow
3}}{d\Omega_{13}}\right)^{(--+++)}_{FnSplit} =\frac{\alpha^2 N_c^2}{E^2}
 \left(\frac{\mu^2}{s}\right)^{\epsilon}\left(\frac{\mu^2}{Q_f^2}\right)^{\epsilon}
 \frac{\alpha }{\pi}\left\{-\frac{1}{\epsilon}\frac{4(c^2+3)}{(1-c^2)^2}
\log\frac{1-\delta}{\delta} \right\};
\end{eqnarray}
  \item
The initial and final splitting for the anti-MHV amplitude
\begin{eqnarray}
&&\hspace{-1.8cm}\left(\frac{d\sigma_{2 \rightarrow
3}}{d\Omega_{13}}\right)^{(--++-)}_{InSplit} =\frac{\alpha^2 N_c^2}{E^2}
 \left(\frac{\mu^2}{s}\right)^{\epsilon}\hspace{-0.1cm}
 \left(\frac{\mu^2}{Q_f^2}\right)^{\epsilon}\hspace{-0.1cm}
 \frac{\alpha}{\pi}\left\{
\frac{1}{\epsilon}\left[\frac{8(c^2+3)}{(1-c^2)^2} \log\delta - \frac{64 (12
c^2+17)}{3(1-c^2)^3}
\right.\right.  \nonumber \\
&&\hspace{-1.8cm}-\left.\left. \frac{4\delta}{(1-c^2)^2} \left( \frac 23 (1+c^2) \delta^2
+ (c^2-5) \delta + 2 (c^2+17) \right)+\left(
\frac{4(c^3\!-\!15c^2\!+\!51c\!-\!45)}{(1-c)^2(1+c)^3}\log\frac{1\!-\!c}{2}
\right. \right.\right.\nonumber \\
&&\hspace{-1.8cm}+ \left.\left.  \left. \frac{8(c^2-6c+21)}{(1-c)(1+c)^3}
\log\frac{1+\delta-c(1-\delta)}{2}+
\frac{32(c^2-4c+7)}{(1+c)^3(1-c)(1+\delta-c(1-\delta))}
\right. \right. \right.\\
 && \hspace{-1.8cm}- \left. \left.\left.
\frac{32(2-c)}{(1\!+\!c)^3(1\!+\!\delta\!-\!c(1\!-\!\delta))^2} \!+\!
\frac{64(1-c)}{3(1\!+\!c)^3(1\!+\!\delta\!-\!c(1\!-\!\delta))^3}\!+\! (c\leftrightarrow
-c) \right) \right]+ \mbox{Finite part}\right\},\nonumber
\end{eqnarray}
\begin{eqnarray}\hspace*{-1.4cm}
\left(\frac{d\sigma_{2 \rightarrow
3}}{d\Omega_{13}}\right)^{(--++-)}_{FnSplit}\hspace{-0.5cm} &=&\!\!\frac{\alpha^2
N_c^2}{E^2}\!
 \left(\frac{\mu^2}{s}\right)^{\epsilon}\hspace{-0.2cm}
 \left(\frac{\mu^2}{Q_f^2}\right)^{\epsilon}\hspace{-0.2cm}
 \frac{\alpha }{\pi} \left\{\frac{1}{\epsilon}\frac{4(c^2+3)}{(1-c^2)^2}\left[
\log\delta\!-\!\delta(\frac 13\delta^2\!-\!\frac 32\delta\!+\!3)\right]\right\};
\end{eqnarray}
  \item
The initial splitting for the fermion  final states ($\delta=1$)
\begin{eqnarray}
&&\hspace{-0.5cm} \left(\frac{d\sigma_{2 \rightarrow 3}}{d\Omega_{13}}\right)^{(--+q\bar
q)}_{InSplit} =\frac{\alpha^2 N_c^2}{E^2}
 \left(\frac{\mu^2}{s}\right)^{\epsilon}\left(\frac{\mu^2}{Q_f^2}\right)^{\epsilon}
 \frac{\alpha }{\pi}\left\{
\frac{16}{\epsilon}\left[\frac{(79+25c^2)}{3(1-c^2)^2}\right.\right.
\\ \nonumber &&\hspace{-0.5cm}\left.\left.+
\frac{2(3-c)^2}{(1-c)(1+c)^3}\log(\frac{1-c}{2})
+\frac{2(3+c)^2}{(1-c)^3(1+c)}\log(\frac{1+c}{2})\right]+\mbox{Finite part}\right\};
\end{eqnarray}
  \item
The initial splitting for the scalar final states ($\delta=1$)
\begin{eqnarray} &&\hspace{-0.5cm}
\left(\frac{d\sigma_{2 \rightarrow
3}}{d\Omega_{13}}\right)^{(--+\Lambda\Lambda)}_{InSplit} =\frac{\alpha^2 N_c^2}{E^2}
 \left(\frac{\mu^2}{s}\right)^{\epsilon}\left(\frac{\mu^2}{Q_f^2}
 \right)^{\epsilon}\frac{\alpha }{\pi}\left\{
\frac{8}{\epsilon}\left[-\frac{2(10+7c^2)}{(1-c^2)^2}\right.\right. \\
&& \nonumber \hspace{-0.5cm}\left.\left.-
\frac{3(5-c)}{(1+c)^3}\log(\frac{1-c}{2})-\frac{3(5+c)}{(1-c)^3}
\log(\frac{1+c}{2})\right] +\mbox{Finite part}\right\}.
\end{eqnarray}
\end{enumerate}

\section{IR Safe Observables in ${\cal N}=4$ SYM}\label{s5}

In the NLO there are  two sets of amplitudes, namely,  the MHV and anti-MHV amplitudes
which contribute to the observables.  The leading order  4-gluon amplitude is both MHV
and anti-MHV and we split it into two parts.  Then one can construct three types of
infrared finite quantities in the NLO of perturbation theory, namely,
\begin{itemize}
\item pure gluonic MHV amplitude
\begin{equation} \label{fin1} \hspace{-1cm}A^{MHV}=\frac
12\left(\frac{d\sigma_{2 \rightarrow
2}}{d\Omega_{13}}\right)^{(--++)}_{Virt}\hspace{-0.3cm}
+\left(\frac{d\sigma_{2 \rightarrow
3}}{d\Omega_{13}}\right)^{(--+++)}_{Real}\hspace{-0.3cm}+
\left(\frac{d\sigma_{2 \rightarrow
2}}{d\Omega_{13}}\right)^{(--+++)}_{InSplit}\hspace{-0.3cm}+
\left(\frac{d\sigma_{2 \rightarrow
2}}{d\Omega_{13}}\right)^{(--+++)}_{FnSplit};\end{equation}
\item pure gluonic anti-MHV amplitude
\begin{equation}  \label{fin2} \hspace{-1cm}B^{anti-MHV}=\frac
12\left(\frac{d\sigma_{2 \rightarrow
2}}{d\Omega_{13}}\right)^{(--++)}_{Virt}\hspace{-0.3cm}
+\left(\frac{d\sigma_{2 \rightarrow
3}}{d\Omega_{13}}\right)^{(--++-)}_{Real}\hspace{-0.3cm}+
\left(\frac{d\sigma_{2 \rightarrow
2}}{d\Omega_{13}}\right)^{(--++-)}_{InSplit}\hspace{-0.3cm}+
\left(\frac{d\sigma_{2 \rightarrow
2}}{d\Omega_{13}}\right)^{(--++-)}_{FnSplit};\end{equation}
\item anti-MHV amplitude with fermions or scalars forming the full
$\mathcal{N}=4$ supermultiplet
\begin{equation} \label{fin3}  \hspace{-1cm}C^{Matter}=
\left(\frac{d\sigma_{2 \rightarrow 3}}{d\Omega_{13}}\right)^{(--+,\ q{\bar
q}+\Lambda\Lambda)}_{Real}+ \left(\frac{d\sigma_{2 \rightarrow
2}}{d\Omega_{13}}\right)^{(--+,\ q{\bar q}+\Lambda\Lambda)}_{InSplit} .
\end{equation}
\end{itemize}
We would like to stress once more  that in each expression
(\ref{fin1},\ref{fin2},\ref{fin3}) \textit{all IR divergencies cancel} for arbitrary
$\delta$ and only the finite part is left.

Two comments are in order. First, this decomposition is valid only
in the leading order in $\alpha$. In the next orders the inclusive
cross-section requires extra emitted particles that takes us away
from the class of the MHV amplitudes. It is not clear whether in
this case one has separate IR-safe sets or everything is mixed
together and only the total cross-section is finite. In the latter case one probably faces the complication
since the non-MHV amplitudes are not known to possess a simple
structure as the MHV ones,  though the origin of this simplicity is unclear.
The second comment concerns the contribution of the matter
fields.  In the leading order we singled it out in the class $C$.
At the same time, in general, there is the contribution of them to the
 virtual part and to the splitting one via the gluon distribution
function. They contain the $1/\epsilon$ terms.
However, the matter field contribution to the virtual part
is proportional to the tree-level $2\times 2$ cross-section with the coefficient
$(\frac 23 n_f+\frac 16 n_s)$ \cite{KTT} and the contribution to the splitting function
comes with the $\beta$-function, i.e. with the same coefficient but with the opposite sign.
Thus,  these contributions have the same structure and completely cancel each
other.   So, in the leading order our separation becomes possible.

Defining now the physical condition for the observation we get
several infrared-safe inclusive cross-sections
\begin{itemize}
\item Registration of \textit{two fastest} gluons of positive helicity
\begin{equation} \label{Fin1} A^{MHV}\Big|_{\delta=1/3}+B^{anti-MHV}\Big|_{\delta=1};
\end{equation}
\item Registration of \textit{one fastest} gluon of positive helicity
\begin{equation} \label{Fin2} A^{MHV}\Big|_{\delta=1/3}+B^{anti-MHV}\Big|_{\delta=1/3}
+C^{Matter}\Big|_{\delta=1};
\end{equation}
\item
Anti-MHV cross-section
\begin{equation} \label{Fin3} B^{anti-MHV}\Big|_{\delta=1}+C^{Matter}\Big|_{\delta=1}.
\end{equation}
\end{itemize}

Relative simplicity of the virtual contribution (\ref{1loop4gluon})
> which contains logarithms and no other special functions suggests a
> similar structure of the real part. However, this is not the case. While the singular terms are simple enough and cancel
completely, the finite parts are usually cumbersome and contain
polylogarithms. The only expression where they cancel corresponds to
the $\delta=1$ case which is possible only for the last set of
observables, namely, for the anti-MHV cross-section (\ref{Fin3}).
Choosing the factorization scale to be $Q_f=E$ we get:
\begin{eqnarray}
&&\hspace{-0.7cm}\left(\frac{d\sigma}{d\Omega_{13}}\right)_{anti-MHV}
=\frac{4\alpha^2 N_c^2}{E^2}\left\{\frac{3+c^2}{(1-c^2)^2}\right.  \\
&& \makebox[2em]{} \hspace{-1.2cm}\left.- \frac{\alpha }{4\pi}\left[ 2\frac{ (c^4\!+\!2
c^3\!+\!4 c^2\!+\!6 c\!+\!19) \log^2(\frac{1-c}{2})}{(1-c)^2 (1+c)^4}
 +2\frac{ (c^4\!-\!2 c^3\!+\!4 c^2\!-\!6 c\!+\!19)
\log^2(\frac{1+c}{2})}{(1-c)^4 (1+c)^2}\right.\right.  \nonumber \\
&& \makebox[2em]{} \hspace{-0.5cm}\left.\left. -8 \frac{(c^2+1)
\log(\frac{1+c}{2})\log(\frac{1-c}{2})}{(1-c^2)^2}
+\frac{6\pi^2(3c^2+13)-5 (61 c^2+99)}{9 (1-c^2)^2}\right.\right. \nonumber \\
&& \makebox[2em]{} \hspace{-0.5cm}\left.\left.-2\frac{(11 c^3\!-\!31 c^2\!-47 c\!-\!133)
\log(\frac{1-c}{2})}{3(1+c)^3 (1-c)^2} +2\frac{(11 c^3\!+\!31 c^2\!-\!47 c\!+\!133)
\log(\frac{1+c}{2})}{3(1-c)^3 (1+c)^2}\right]\right\}.\nonumber
\end{eqnarray}

One can see that even this expression does not repeat the  form of the Born amplitude and
does not have any simple structure. While the dependence on the  parameter $\mu$ of
dimensional reduction is completely canceled, the finite answer, as in the toy model
example, depends on the factorization scale. This dependence comes from the asymptotic
states which violate conformal invariance of the Lagrangian. This dependence seems to be
unavoidable  and reflects the act of measurement. Construction of observables which do
not contain any external scale remains an open question.

\section{Discussion}\label{s6}

Remarkable factorization properties of the MHV amplitudes accumulated in the BDS ansatz
(with the so far unknown modification) and duality with the string amplitudes via the
AdS/CFT correspondence seem to suggest the way to get the exact solution of the
$\mathcal{N}=4$ SYM theory. However, "to solve the model" might have a different meaning.
Calculation of divergences and understanding of their structure is very useful but surely
not enough, it is the finite part that we are really for. The knowledge of the S-matrix
would be the final goal though the definition of the S-matrix in conformal theory is a
problem. Even in the absence of the UV divergences there are severe IR problems and
matrix elements do not exist after removal of regularization.

The purpose of this paper is to present all the details of the calculation with explicit
cancellation  of the infrared divergencies in properly defined cross-sections in the
planar limit for $\mathcal{N}=4$ SYM. The main results were summarized in our short
letter~\cite{we}. We do obtain IR safe observables in the weak coupling regime in the
next-to-leading order of PT which are calculated analytically. The same procedure can
also be  applied to $\mathcal{N}=8$ SUGRA \cite{toappear}.

Unfortunately, our calculation has demonstrated that the simple structure of the
amplitudes governed by the cusp anomalous dimension has been totally washed out by
complexity of the real emission matrix elements integrated over the phase space. This
means that either the $\mathcal{N}=4$ SYM theory  does not allow such a simple
factorizable solution or that we considered the unappropriate observables that do not
bear the impact of the $\mathcal{N}=4$ symmetry. One can obviously see the presence of
$\mathcal{N}=4$ supermultiplet in the virtual part but not in the real emission. It would
be  of great importance to find such quantities.

Another unfortunate feature of inclusive cross-sections is the dependence on the
factorization scale. The experience of QCD, which is very similar to the $\mathcal{N}=4$
SYM theory from the point of view of the IR problems, tells us that in inclusive
cross-sections the IR divergences cancel and one has finite physical observables.
However, in QCD one has confinement and considers the scattering of the bound states
(hadrons, glueballs) rather than the individual particles. In this case, one usually
factorizes  the hard part from the soft part introducing the factorization scale. The
dependence on this scale is canceled between the hard and soft parts contrary to our case
where only the hard part is present. But in QCD one also has an additional scale. The
parton distributions  are defined experimentally at some scale $Q_0$ and the dependence
on this scale is left. This dependence is governed by the same DGLAP equations as the
dependence on the factorization scale, so from this point of view the situation in QCD is
not better than in our case.

In both the cases one has to introduce some parton distributions which are the functions
of a fraction of momenta and, in higher orders, of momenta transferred. This leads to the
appearance of an additional scale which breaks the conformal invariance. One might think
of some observables where this scale dependence is canceled, like the ratio of some
cross-sections, etc. We have not found such quantities so far, though the construction of
such truly conformal observables is of great interest. Probably, they might have the
desired simple structure.

There is an interesting duality between the MHV amplitudes and the Wilson loop, between
the weak and the strong coupling regime~\cite{Dks,am1,Dhks1}. Perhaps, it would be
possible, using the AdS/CFT correspondence, to construct the IR safe observables in the
strong coupling limit (similarly to what we did here) and to shed some light on the  true
calculable objects in conformal theories.

\section*{Acknowledgements}

We would like to thank N. Beisert, A. Gorsky, A. Grozin,  G. Korchemsky, D. Kosower, A.
Kotikov, L. Lipatov, T. McLoughlin, S. Mikhailov, R. Roiban, A. Slavnov and V. Smirnov
for valuable discussions. Financial support from RFBR grant \# 08-02-00856 and grant of
the Ministry of Education and Science of the Russian Federation \# 1027.2008.2 is kindly
acknowledged. Two of us LB and GV are partially supported by the Dynasty Foundation. GV
is grateful to BLTP, JINR where  most of the work was carried out. DK acknowledges the
hospitality of  KEK, Tsukuba where the paper was finished.

\renewcommand{\thesection}{}
\section{\hspace*{-0.6cm}Appendix A. Computation of partial amplitudes}
\setcounter{equation}0
\renewcommand{\theequation}{A.\arabic{equation}}
To calculate the cross-section we need the squared matrix elements summed over helicities
and color. They can be expressed in terms of  the corresponding partial amplitudes
\cite{Parke}
\begin{eqnarray}\label{hel}
&&|\mathcal{M}_n(p_1,...,p_n)|^2=g^{2n-4}(\frac{g^2N_c}{16\pi^2})^{2l}\sum_{colors}
|\mathcal{A}_n^{(l-loop)}|\nonumber\\
&&=g^{2n-4}N_c^{n-2}(N_c^2-1)(\frac{g^2N_c}{16\pi^2})^{2l}\sum_{\sigma \in
P_{n-1}}|A_n^{(l-loop)}(p_1,p_{\sigma(2)},...,p_{\sigma(n)})|^2.
\end{eqnarray}

For the massless  partial helicity amplitudes it is convenient to use the so-called
spinor helicity formalism   initially  introduced in \cite{sp1,sp2,sp3} (for a review see
\cite{Dixon}). In this formalism the on-shell momenta of every $i$-th external massless
particle $p^{(i)}_{\mu}p^{(i)\mu}=0$ is represented in terms of a pair of massless
commuting spinors $\lambda_{a}^{(i)}$ and $\bar{\lambda}_{\dot{a}}^{(i)}$ of positive and
negative chirality in the following way:
\begin{eqnarray}
p^{(i)}_{\mu} \longrightarrow
p^{(i)}_{a\dot{a}}=p^{(i)}_{\mu}(\sigma^{\mu})_{a\dot{a}}=
\lambda_{a}^{(i)}\bar{\lambda}_{\dot{a}}^{(i)}.
\end{eqnarray}
The spinor inner product is defined by:
\begin{eqnarray}
\epsilon^{ab}\lambda_{a}^{(i)}\lambda_{b}^{(j)}=\langle\lambda^{(i)}\lambda^{(j)}\rangle
\doteq \langle ij\rangle,
~~~\epsilon^{\dot{a}\dot{b}}\bar{\lambda}_{\dot{a}}^{(i)}\bar{\lambda}_{\dot{b}}^{(j)}
=[\bar{\lambda}^{(i)}\bar{\lambda}^{(j)}] \doteq [ ij],
\end{eqnarray}
thus the complex conjugation of the product is
\begin{equation}
(\langle ij\rangle)^{*}=[ ij].
\end{equation}
The scalar product of the two light-like momenta can be represented in terms of these
products as
\begin{equation}
p^{\mu(i)}p^{(j)}_{\mu}=\frac{1}{2}\langle ij\rangle[ ij],
\end{equation}
or equivalently
\begin{equation}
\langle ij\rangle[ij]=s_{ij},
\end{equation}
where the standard notation $(p_{i}+p_{j})^2=s_{ij}$ is used.

All the tree-level partial MHV amplitudes  can be combined into a single $\mathcal{N}=4$
supersymmetric expression, first suggested by Nair \cite{Nair}:
\begin{eqnarray}
\mathcal{Z}_{n}^{\mathcal{N}=4~MHV}=\delta^{8}\left(\sum_{i=1}^{n}\lambda_{i}^{a}\eta_{i}^{(A)}\right)
\frac{1}{\prod_{i=1}^{n}\langle i,i+1\rangle}.
\end{eqnarray}
where $\eta_{i}^{(A)}$ are the Grassmannian coordinates, $A=1,...,4$ is the $SU(4)_R$
fundamental representation index. $\mathcal{Z}_{n}^{\mathcal{N}=4~MHV}$ is invariant
under $SU(4)_R$ transformations of $\eta_{i}^{(A)}$ and under the cyclic permutations of
momentum labels $i$. In the product $\prod_{i=1}^{n}\langle i,i+1\rangle$ one has to
identify $i+n$ with $i$. The Grassmannian-valued delta function is defined in the usual
way:
\begin{equation}
\delta^{8}\left(\sum_{i=1}^{n}\lambda_{i}^{a}\eta_{i}^{(A)}\right)=
\prod_{A=1}^{4}\frac{1}{2}\left(\sum_{i=1}^{n}\lambda_{i}^{a}\eta_{i}^{(A)} \right)
\left( \sum_{k=1}^{n}\lambda_{ka}\eta_{k}^{(A)}\right)
 =\frac{1}{16}\prod_{A=1}^{4}\sum_{i,k=1}^{n}\langle ik
\rangle \left(\eta_{i}^{(A)}\eta_{k}^{(A)}\right)
\end{equation}
So one can rewrite $\mathcal{Z}_{n}^{\mathcal{N}=4~MHV}$ as
\begin{eqnarray}
\mathcal{Z}_{n}^{\mathcal{N}=4~MHV}=\frac{1}{16}\sum_{i,...,c=1}^{n}\langle
ik \rangle \langle lm \rangle\langle ab \rangle\langle dc \rangle
\left(\eta_{i}^{(1)}\eta_{k}^{(1)}\eta_{l}^{(2)}\eta_{m}^{(2)}\eta_{a}^{(3)}\eta_{b}^{(3)}
\eta_{d}^{(4)}\eta_{c}^{(4)}\right)\frac{1}{\mathcal{P}_{n}},
\end{eqnarray}
where
\begin{equation}
\mathcal{P}_{n}=\prod_{i=1}^{n}\langle i,i+1\rangle.
\end{equation}

Using the Taylor expansion of $\mathcal{Z}_{n}^{\mathcal{N}=4~MHV}$ in powers of
$\eta^{(A)}$ one gets the sum of $(\frac{n(n-1)}{2})^{4}$ terms each involving a product
of 8 distinct $\eta_{i}^{(A)}$. One can identify the coefficient of the product of 8
$\eta$'s in each term in the expansion with a particular tree component partial
amplitude. It is very useful to define the following differential operators with the
self-explanatorily notation:
\begin{eqnarray}
\hat{g}^{+}(i)&=&1, \nonumber \\
\hat{g}^{-}(i)&=&\frac{1}{4!}\epsilon^{ABCD}\frac{\partial^{4}}{\partial\eta_{i}^{(A)}
\partial\eta_{i}^{(B)}
\partial\eta_{i}^{(C)}\partial\eta_{i}^{(D)}}
=
\frac{\partial^{4}}{\partial\eta_{i}^{(1)}\partial\eta_{i}^{(2)}\partial\eta_{i}^{(3)}
\partial\eta_{i}^{(4)}},\nonumber\\
\hat{q}^{+}(i)^{A}&=&\frac{\partial}{\partial\eta_{i}^{(A)}},\\
\nonumber
\hat{q}^{-}(i)_{A}&=&-\frac{1}{3!}\epsilon_{ABCD}\frac{\partial^{3}}{\partial\eta_{i}^{(B)}
\partial\eta_{i}^{(C)}\partial\eta_{i}^{(D)}},\\ \nonumber
\hat{\Lambda}(i)^{AB}&=&\frac{\partial^{2}}{\partial\eta_{i}^{(A)}\partial\eta_{i}^{(B)}},\\
\nonumber
\hat{\Lambda}(i)_{CD}&=&\frac{1}{2!}\epsilon^{ABCD}\frac{\partial^{2}}{\partial\eta_{i}^{(A)}\partial\eta_{i}^{(B)}}.\nonumber
\end{eqnarray}
Taking various combinations of products of these operators one can construct a set of
8-th order differential operators. These 8-th order differential operators act as
projectors on the component partial amplitudes: $\hat{q}^{+}(i)^{A}$ corresponds to the
fermion $q^{A,+}$  of the ${\cal N}=4$ supermultiplet, $\hat{q}^{-}_{A}$ to
$\bar{q}_A^-$, $\hat{\Lambda}^{AB}(i)$  to $\Lambda^{AB}$, and $\hat{\Lambda}_{AB}(i)$ to
$\Lambda_{AB}$.

For example, the Parke-Taylor $n$-gluon amplitude can be written as:
\begin{equation}
A_{n}^{(tree)}(g^{-}g^{-}g^{+}...g^{+}) =
\hat{g}^{-}(1)\hat{g}^{-}(2)\hat{g}^{+}(3)...\hat{g}^{+}(n)\mathcal{Z}_{n}^{\mathcal{N}=4~MHV}=\langle
12\rangle^{4}\frac{1}{\mathcal{P}_{n}},
\end{equation}
and the squared partial amplitude $|A_{n}^{(tree)}(g^{-}g^{-}g^{+}...g^{+})|^2$ then
takes the simple form (it is implemented that momenta  are ordered as
$p_1,p_2,p_3,...,p_n$)
\begin{eqnarray}
|A_{n}^{(tree)}(g^{-}g^{-}g^{+}...g^{+})|^2&=&
A_{n}^{(tree)}(g^{-}g^{-}g^{+}...g^{+})A_{n}^{(tree)}(g^{-}g^{-}g^{+}...g^{+})^{*}\nonumber\\
&=&\frac{\langle 12\rangle^{4}[
12]^{4}}{\mathcal{P}_{n}\mathcal{P}_{n}^{*}}=\frac{s_{12}^4}{s_{12}s_{23}...s_{n1}}.\label{AA}
\end{eqnarray}
To extract from (\ref{hel}) some specific helicity configuration for the MHV amplitude,
one has to sum over the permutations only in the denominator of (\ref{AA})
\cite{Mangano}. So, for example, for the Parke-Taylor $n$-gluon amplitude one has
\begin{eqnarray}
|M_{n}^{(tree)}(g^{-}g^{-}g^{+}...g^{+})|^2=g^{2n-4}N_c^{n}\ s_{12}^4\sum_{\sigma \in
P_{n-1}}\frac{1}{s_{1\sigma(2)}s_{\sigma(2)\sigma(3)}...s_{\sigma(n)\sigma(1)}}.
\end{eqnarray}

The anti-MHV amplitudes also needed for our computation can be obtained from the
corresponding conjugated MHV amplitudes. For example the anti-MHV amplitude
$A_{5}(g^{-}g^{-}g^{+}g^{-}g^{+})$ can be obtained from the MHV amplitude
$A_{5}(g^{+}g^{+}g^{-}g^{+}g^{-})$ by making a complex conjugation.

Below we present the list of four- and five-point tree amplitudes which are relevant to
our calculation. The four-point amplitudes are
\begin{eqnarray}
&&\hspace*{-1cm}A_{4}^{(tree)}(g^{-}g^{-}g^{+}g^{+}) =
\hat{g}^{-}(1)\hat{g}^{-}(2)\hat{g}^{+}(3)\hat{g}^{+}(4)
\mathcal{Z}_{4}^{\mathcal{N}=4~MHV}
=\langle 12\rangle^{4}\frac{1}{\mathcal{P}_{4}},\\
&&\hspace*{-1cm}A_{4}^{(tree)}(g^{-}g^{+}g^{-}g^{+})=g^{-}(1)g^{+}(2)g^{-}(3)g^{+}(4)
\mathcal{Z}_{4}^{\mathcal{N}=4~MHV} = \langle 13\rangle^{4}\frac{1}{\mathcal{P}_{4}},\\
&&\hspace*{-1cm}A_{4}^{(tree)}(g^{-}q^{A}g^{+}\bar{q}_{A}) =
\hat{g}^{-}(1)\hat{q}^{A,+}(2)\hat{g}^{+}(3)\hat{q}^{-}_{A}(4)
\mathcal{Z}_{4}^{\mathcal{N}=4~MHV}
=\langle12\rangle\langle14\rangle^3 \frac{1}{\mathcal{P}_4},\\
&&\hspace*{-1cm}A_{4}^{(tree)}(g^{-}\bar{q}_{A}g^{+}q^{A}) =
\hat{g}^{-}(1)\hat{q}^{-}_{A}(2)\hat{g}^{+}(3)\hat{q}^{A,+}(4)
\mathcal{Z}_{4}^{\mathcal{N}=4~MHV}
=\langle14\rangle\langle12\rangle^3 \frac{1}{\mathcal{P}_4},\\
&&\hspace*{-1cm}A^{(tree)}_{4}(g^{-}\Lambda^{\tt AB}g^{+}\Lambda_{\tt AB})=
\hat{g}^{-}(1)\hat{\Lambda}^{\tt AB}(2)\hat{g}^{+}(3)\hat{\Lambda}_{\tt AB}(4)
\mathcal{Z}_{4}^{\mathcal{N}=4~MHV}
=\frac{\langle12\rangle^2\langle14\rangle^2}{\mathcal{P}_4}.
\end{eqnarray}
For the computation  of the  real emission  we need the five-point tree amplitudes
\begin{eqnarray}
&&\hspace*{-1.7cm}A_{5}^{(tree)}(g^{-}g^{-}g^{+}g^{+}g^{+})=
\hat{g}^{-}(1)\hat{g}^{-}(2)\hat{g}^{+}(3)\hat{g}^{+}(4)\hat{g}^{+}(5)
\mathcal{Z}_{5}^{\mathcal{N}=4~MHV} = \langle 12\rangle^{4}\frac{1}{\mathcal{P}_{5}},\\
&&\hspace*{-1.7cm}A_{5}^{(tree)}(g^{-}g^{-}g^{+}g^{-}g^{+})=
(\hat{g}^{+}(1)\hat{g}^{+}(2)\hat{g}^{-}(3)\hat{g}^{+}(4)\hat{g}^{-}(5)
\mathcal{Z}_{5}^{\mathcal{N}=4~MHV})^{\ast} = [35]^{4}\frac{1}{\mathcal{P}_{5}^{\ast}},\\
&&\hspace*{-1.7cm}A_{5}^{(tree)}(g^{-}g^{-}g^{+}g^{+}g^{-})=
(\hat{g}^{+}(1)\hat{g}^{+}(2)\hat{g}^{-}(3)\hat{g}^{-}(4)\hat{g}^{+}(5)
\mathcal{Z}_{5}^{\mathcal{N}=4~MHV})^{\ast} = [35]^{4}\frac{1}{\mathcal{P}_{5}^{\ast}},\\
&&\hspace*{-1.7cm}A_{5}^{(tree)}(g^{-}g^{-}g^{+}q^{A}\bar{q}_{A})=
(\hat{g}^{+}(1)\hat{g}^{+}(2)\hat{g}^{-}(3)\hat{q}^{A,+}(4)\hat{q}_{A}^-(5)
\mathcal{Z}_{5}^{\mathcal{N}=4~MHV})^{\ast} = \frac{[34]^{3}[35]}{\mathcal{P}_{5}^{\ast}},\\
&&\hspace*{-1.7cm}A_{5}^{(tree)}(g^{-}g^{-}g^{+}\bar{q}_{A}q^{A})=(\hat{g}^{+}(1)\hat{g}^{+}(2)
\hat{g}^{-}(3)\hat{q}_{A}^-(4)\hat{q}^{A,+}(5)\mathcal{Z}_{5}^{\mathcal{N}=4~MHV})^{\ast}
=
\frac{[34][35]^{3}}{\mathcal{P}_{5}^{\ast}},\\
&&\hspace*{-1.7cm}A_{5}^{(tree)}(g^{-}g^{-}g^{+}\Lambda^{\tt AB}\Lambda_{\tt AB})=
(\hat{g}^{+}(1)\hat{g}^{+}(2)\hat{g}^{-}(3)\hat{\Lambda}^{\tt AB}(4) \hat{\Lambda}_{\tt
AB}(5) \mathcal{Z}_{5}^{\mathcal{N}=4~MHV})^{\ast}
=\frac{[34]^{2}[35]^{2}}{\mathcal{P}_{5}^{\ast}}.
\end{eqnarray}

We also provide the list of the Born cross-sections used in Sect.4.
\begin{eqnarray}
\left(\frac{d\sigma_{2 \rightarrow 2}}{d\Omega_{13}}\right)_{(tree)}^{(--++)}
&=&\frac{\alpha^2N_c^2}{2E^2}
s^2 \left(\frac{s^2+t^2+u^2}{t^2u^2}\right)\left(\frac{\mu^2}{s}\right)^\epsilon,
\label{A25}\\
\left(\frac{d\sigma_{2 \rightarrow 2}}{d\Omega_{13}}\right)_{(tree)}^{(-+-+)}
&=&\frac{\alpha^2N_c^2}{2E^2}
t^2 \left(\frac{s^2+t^2+u^2}{s^2u^2}\right)\left(\frac{\mu^2}{s}\right)^\epsilon,\\
\left(\frac{d\sigma_{2 \rightarrow 2}}{d\Omega_{13}}\right)_{(tree)}^{(-++-)}
&=&\frac{\alpha^2N_c^2}{2E^2}
u^2 \left(\frac{s^2+t^2+u^2}{t^2s^2}\right)\left(\frac{\mu^2}{s}\right)^\epsilon,\\
\left(\frac{d\sigma_{2 \rightarrow 2}}{d\Omega_{13}}\right)_{(tree)}^{(-q+\bar q)}
&=&\frac{\alpha^2N_c^2}{2E^2}
|u| \left(\frac{s^2+t^2+u^2}{t^2s}\right)\left(\frac{\mu^2}{s}\right)^\epsilon,\\
\left(\frac{d\sigma_{2 \rightarrow 2}}{d\Omega_{13}}\right)_{(tree)}^{(-\bar q+ q)}
&=&\frac{\alpha^2N_c^2}{2E^2}
s \left(\frac{s^2+t^2+u^2}{t^2|u|}\right)\left(\frac{\mu^2}{s}\right)^\epsilon,\\
\left(\frac{d\sigma_{2 \rightarrow 2}}{d\Omega_{13}}\right)_{(tree)}^{(-\Lambda+\Lambda)}
&=&\frac{\alpha^2N_c^2}{2E^2}
 \left(\frac{s^2+t^2+u^2}{t^2}\right)\left(\frac{\mu^2}{s}\right)^\epsilon.\label{A30}
\end{eqnarray}
These cross-sections are written down for the set of momenta ($p_1,p_2,p_3,p_4$) with the
conservation law $p_1+p_2=p_3+p_4$. In the case of initial splitting, according to
(\ref{AP}),  one should use the cross-sections calculated for the set
($zp_1,p_2,p_3,p_4$) with a new conservation law $zp_1+p_2=p_3+p_4$. To  get them, one
should substitute the modified values for the Mandelstam variables (\ref{KinSpl}) into
(\ref{A25}-\ref{A30}) and multiply the cross-sections by the factor $4/(1+z-c(1-z))^2$
which comes from the modified delta function $\delta^D(zp_1+p_2-p_3-p_4)$. The same
procedure but with the replacement $c\leftrightarrow -c$ refers to the $p_1
\leftrightarrow p_2$ case.

\section{\hspace*{-0.6cm}Appendix B. Calculation of phase space integrals}
\renewcommand{\thesubsection}{B.\arabic{subsection}}
\setcounter{equation}0
\renewcommand{\theequation}{B.\arabic{equation}}

Consider the structure of the matrix elements in detail. First off all it is convenient
to rewrite the standard three-particle phase space
\begin{equation}
d \phi_{3} =  \delta^{+}(p_3^2)\frac{d^{D}p_{3}}{(2\pi)^{D-1}}
\delta^{+}(p_{4}^2)\frac{d^{D} p_{4}}{(2\pi)^{D-1}} \delta^{+}(p_{5}^2)\frac{d^{D}
p_{5}}{(2\pi)^{D-1}} (2\pi)^D\delta^{D}(p_{1}+p_{2}-p_{3}-p_{4}-p_{5})
\end{equation}
in the following form:
\begin{equation}
d \phi_{3} = \delta^{+}(p_3^2)\frac{d^{D}p_{3}}{(2\pi)^{D-1}} \delta^{+}((p_{4}-k)^2)
\frac{d^{D} p_{4}}{(2\pi)^{D-1}} \delta^{+}(k^2) \frac{d^{D} k}{(2\pi)^{D-1}}
 (2\pi)^D\delta^{D}(p_{1}+p_{2}-p_{3}-p_{4}).
\end{equation}
The integral we are interested in is
\begin{equation}\label{INT}
\int |\mathcal{M}_5|^2 \mathcal{S}_3(p_3,k,p_4-k) d\phi_3,
\end{equation}
where the matrix element $|\mathcal{M}_5|^2$  for the five-point amplitude consists of 12
terms
with identical numerator  but different denominators. The typical integrand looks like
\begin{equation}\label{kint}
I = \frac{2s_{12}^4}{s_{13}s_{25}s_{35}s_{24}s_{14}} =
\frac{2((p_1+p_2)^2)^4}{(p_1-p_3)^2(p_2-k)^2(-p_3-k)^2(p_2-[p_4-k])^2(p_1-[p_4-k])^2}.
\end{equation}

Our strategy is to use the on-shell conditions to simplify all the terms in the sum so
that the integral over $d^Dk$ can be calculated exactly. For the remaining integrals we
evaluate the necessary terms of the $\epsilon$-expansion.

Taking into account  the conservation of the momentum $p_1 + p_2=p_3 + p_4$ and the
on-shell conditions
$$p_1^2=0, \quad p_2^2=0,\quad p_3^2=0,\quad k^2=0,\quad (p_4-k)^2 =0$$
one can rewrite the integrand (\ref{kint}) as
\begin{equation}
I = \frac{(p_1,p_2)^4}{(p_1,p_3)(p_2,k)(p_3,k)(p_2,p_4-k)(p_1,p_4-k)},
\end{equation}
where we use the notation (p,k)=pk for the scalar product.

The next step is to use the partial fraction with respect to $k$
\begin{eqnarray}
I &=& \frac{(p_1 , p_2)^4}{(p_1 , p_3)(p_3 , k)(p_1, p_4-k)} \frac{1}{(p_2,p_4)}
\left(  \frac{1}{(p_2,k)} + \frac{1}{(p_2,p_4-k)} \right) \nonumber \\
&=& \frac{(p_1,p_2)^4}{(p_1,p_3)(p_2,p_4)} \frac{1}{(p_3,k)(p_2,k)(p_1,p_4\!-\!k)} +
\frac{(p_1,p_2)^4}{(p_1,p_3)(p_2,p_4)} \frac{1}{(p_3,k)(p_2,p_4\!-\!k)(p_1, p_4\!-\!k)} \nonumber
\\ &=& \frac{(p_1,p_2)^4}{(p_1,p_3)(p_2,p_4)} \frac{1}{(p_1,p_4)-(p_4,p_4)/2} \nonumber
\\ && \makebox[3em]{} \times \left( \frac{1}{(p_3,k)(p_2,k)} - \frac{1}{(p_3,
k)(p_1,p_4-k)} + \frac{1}{(p_2,k)(p_1,p_4\!-\!k)} \right) \nonumber \\
 && + \frac{(p_1,p_2)^4}{(p_1,p_3)(p_2,p_4)}
\frac{1}{(p_1,p_4)+(p_2,p_4)-(p_4,p_4)/2}\\ && \makebox[3em]{} \times \left(
\frac{1}{(p_3,k)(p_2,p_4-k)} + \frac{1}{(p_3, k)(p_1,p_4-k)} +
\frac{1}{(p_2,p_4\!-\!k)(p_1,p_4\!-\!k)} \right)\nonumber
\end{eqnarray}
so that one gets at most two brackets with momentum $k$ in the denominator.

In the case of momentum $k$ in the numerator, this procedure also works but with some
variation. For example, one has
\begin{eqnarray}
J&=&\frac{(p_1,k)}{(p_1,p_2)(p_1, p_3)(p_3, p_4-k)(p_2 ,k)(k, p_4 - k)}\nonumber
\\ &=& \frac{1}{(p_1 , p_2)(p_1 ,p_3)(p_4^2/2)}
\frac{(p_3,k)+p_4^2/2-(p_2,k)}{(p_2,k)(p_3,p_4-k)} \nonumber \\ &=& \frac{1}{(p_1 ,
p_2)(p_1 ,p_3)(p_4^2/2)} \frac{(p_3,p_4)-(p_3,p_4-k)+p_4^2/2-(p_2,k)}{(p_2,k)(p_3,p_4-k)}
\nonumber \\ &=& \frac{(p_3,p_4)+p_4^2/2}{(p_1 , p_2)(p_1
,p_3)(p_4^2/2)(p_2,k)(p_3,p_4-k)}  \\ && \makebox[2em]{} - \frac{1}{(p_1 , p_2)(p_1
,p_3)(p_4^2/2)(p_2,k)} - \frac{1}{(p_1 , p_2)(p_1 ,p_3)(p_4^2/2)(p_3,p_4-k)}.\nonumber
\end{eqnarray}
Since we usually have $(p_i,k)^4$ in the numerator this procedure has to be applied
several times. This way we increase the number of terms in the integrand  but drastically
simplify the integration.

\begin{figure}[h,t]
\begin{center}
 \leavevmode
  \epsfxsize=4.3cm
 \epsffile{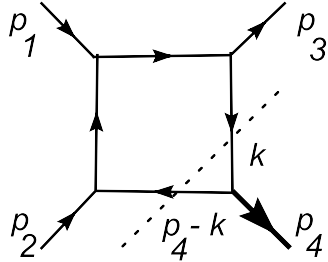}
\caption{The box diagram corresponding to the integral (\ref{exint})\label{box}}
\end{center}
\end{figure}

 The resulting  integrals over $k$ have the standard form
\begin{eqnarray}
\int d^{D} k \delta^{+}((p_{4}-k)^2) \delta^{+}(k^2) \Upsilon_{i}\ ,
\end{eqnarray}
where\vspace{-0.3cm}
\begin{eqnarray}
\Upsilon_{1} &=& \frac{1}{(p_i, k)^a(p_j, k)^b}, \nonumber \\
\Upsilon_{2} &=& \frac{1}{(p_i, k)^a(p_j,p_{4} - k)^b}, \nonumber \\
\Upsilon_{3} &=& \frac{1}{(p_i, p_{4} - k)^a(p_j,p_{4} - k)^b} \nonumber
\end{eqnarray}
and can be calculated by the method of unitarity.  They correspond to the box-type
diagrams and one can perform the cuts and then take the imaginary part. For example, the
integral
\begin{equation}\label{exint}
 \int \frac{d^D k \delta^+(k^2)\delta^+((p_4-k)^2)}{(p_1+k)^2(p_2-p_4+k)^2},
\end{equation}
where $p_1^2=p_2^2=p_3^2=0$ and $p_4^2 \neq 0$ can be obtained from the box-diagram shown
in Fig.\ref{box}.

For the first time this integral was calculated by van Neerven \cite{Neerven} and the
answer is given by
\begin{eqnarray}\label{firstint}
&& \int d^{D} k \delta^{+}((p_{4}-k)^2) \delta^{+}(k^2) \Upsilon_{i}= \frac{(p_4^2/4)^{-
\epsilon}}{(p_i p_4)^a(p_j p_4)^b} \Theta(p_4^2)
\\ \nonumber &\times& 2 \pi \ 2^{-a-b}
\frac{\Gamma(D-3)\Gamma(D/2-1-a)\Gamma(D/2-1-b)}{\Gamma^2(D/2-1)\Gamma(D-2-a-b)}
\phantom{}_2F_1 (a,b;\frac{D}{2}-1|\widetilde{\Upsilon}_{i}),
\end{eqnarray}
where
\begin{eqnarray}
\widetilde{\Upsilon}_{1}& = &1 - \frac{(p_{i},p_{j})(p_{4},p_{4})}{2 (p_i, p_{4})(p_j,
p_{4})}, \nonumber \\
\widetilde{\Upsilon}_{2} &=& \frac{(p_{i},p_{j})(p_{4},p_{4})}{2 (p_i, p_{4})(p_j,
 p_{4})}, \nonumber \\
\widetilde{\Upsilon}_{3} &=& 1 - \frac{(p_{i},p_{j})(p_{4},p_{4})}{2 (p_i, p_{4})(p_j,
p_{4})}. \nonumber
\end{eqnarray}

Removing the integral over $d^Dp_4$ with the help of the delta function we are left with
the last integration over $d^Dp_3$. Using $\delta^{+}(p_3^2)$ one can take the integral
over $p_3^0$ and going to the spherical coordinates
$d^{D-1}\textbf{p}_3=|\textbf{p}_3|^{D-2}d|\textbf{p}_3|d\Omega_{13}$ arrive to the
single integration over the modulus of $|\textbf{p}_3|$.

Here we face the problem of singularity at $|\textbf{p}_3|=0$. It comes from the delta
function in the integration over $p_3^0$ and in some cases is not compensated by the
matrix element. For two matrix elements corresponding to the MHV ($g^+g^+\to g^+g^+g^+$)
and anti-MHV ($g^+g^+\to g^+g^+g^-$) amplitudes, the first case is singular while the
second is not. However, as we explained earlier, we cut the integral over
$|\textbf{p}_3|$ at $(1-\delta)E/2$ and no singularity appears.

Let us now turn to the calculation of the last integral. Since $p_3$ is a dimensionful
parameter, it is appropriate to change variable to dimensionless one using
\begin{equation}\label{change}
|\textbf{p}_3|=\frac{E}{2} (1-x).
\end{equation}
Then the integral over $x$ goes from $0$ to $\delta$.

The typical  integral to be calculated is of the form
\begin{equation}\label{intapp}
\int_{0}^{\delta} d x \frac{x^{\alpha-1} (1-x)^{\beta -1}}{1+px}
\phantom{}_2F_1(1,-\epsilon;1-\epsilon; q x^{m} (1-x)^{n}),
\end{equation}
where $m$ and $n$ take the values
 $$(m,n) = \{(1,0) ,\quad (0,2), \quad (1,-2), \quad (-1,2)\}.$$

For our purposes we need to calculate this integral to the order $\cal{O}(\epsilon)$. The
source of divergence is the singularity at $x=0$. When $\delta \neq 1$, one can expand
the hypergeometric function in $\epsilon$ up to the order $\epsilon$ and then calculate
the integral. Then for the configurations $(1,0)$, $(0,2)$ and $(1,-2)$ the calculation
is straightforward while for the case of $(-1,2)$ one first  makes the transformation of
the argument of the hypergeometric function from $z$ to $1/z$
\begin{eqnarray}\label{transform}
\phantom{}_2F_1(a,b;c|z)&=& \frac{\Gamma (c) \Gamma (-a+b)}{\Gamma (c-a) \Gamma (b)}
\phantom{}_2F_1\left(a,a-c+1;a-b+1|\frac{1}{z}\right) (-z)^{-a} \nonumber \\ &  +&
\frac{\Gamma (c) \Gamma (a-b)}{\Gamma (a) \Gamma (c-b)}
\phantom{}_2F_1\left(b,-c+b+1;-a+b+1|\frac{1}{z}\right) (-z)^{-b}
\end{eqnarray} and then apply the expansion.

For example, consider the integral
\begin{equation}\label{examp}
I_2= \int_{0}^{\delta} d x x^{-1-\epsilon} (1-x)^{-2 \epsilon}
\phantom{}_2F_1(1,-\epsilon;1-\epsilon; q x^{-1} (1-x)^{2}).
\end{equation}
After applying the transformation (\ref{transform}) it is reduced to the following form
\begin{eqnarray}
&&I_2= \int_{0}^{\delta} d x \left( \Gamma (1-\epsilon) \Gamma
(1+\epsilon) \left(\frac{q (1-x)^2}{x}\right)^{\epsilon} \right. \nonumber \\
&& \left. \makebox[3em]{} +\frac{x \Gamma (-1-\epsilon) \Gamma (1-\epsilon)
\phantom{}_2F_1\left(1,1+\epsilon;2+\epsilon;-\frac{x}{q (1-x)^2}\right)}{q (1-x)^2
\Gamma (-\epsilon)^2} \right).
\end{eqnarray}
Performing the expansion over $\epsilon$ one gets to the order of $\cal{O}(\epsilon)$
\begin{eqnarray}
&&I_2= -\frac{1}{2 \epsilon}+\left(\log \delta-\frac{\log q}{2}\right) + \left[-\log ^2
\delta+\log q \log \delta-\log \left((\delta-1)^2 q\right) \log \delta \right. \nonumber
\\ && \left. -\log \left(\frac{2 (\delta-1) q-\sqrt{1-4 q}+1}{2 q}\right) \log
\delta-\log \left(\frac{2 (\delta-1) q+\sqrt{1-4 q}+1}{2 q}\right) \log \delta  \right.
\nonumber \\ && \left. +\log \left(q \delta^2-2 q \delta+\delta+q\right) \log
\delta-\frac{\log ^2 q}{4}
-2 Li_2(1-\delta) -\frac{\pi ^2}{12} \right. \nonumber \\
&& \left. -\log \left(\frac{-2 q+\sqrt{1-4 q}+1}{2 q}\right) \log
\left(-\frac{2 q}{-2 q+\sqrt{1-4 q}+1}\right) \right. \nonumber \\
&& \left. -\log \left(\frac{2 q}{2 q+\sqrt{1-4 q}-1}\right) \log
\left(-\frac{2 q+\sqrt{1-4 q}-1}{2 q}\right) \right. \nonumber \\
\nonumber && \left. +\log \left(\frac{2 \delta q}{2 q+\sqrt{1-4 q}-1}\right) \log
\left(\frac{2 (\delta-1) q-\sqrt{1-4 q}+1}{2 q}\right) \right.
\\ \nonumber && \left. +\log \left(-\frac{2 \delta q}{-2 q+\sqrt{1-4 q}+1}\right)
\log \left(\frac{2 (\delta-1) q+\sqrt{1-4 q}+1}{2 q}\right)\right.
\\ \nonumber && \left. +Li_2\left(\frac{2 (\delta-1) q+\sqrt{1-4
q}+1}{-2 q+\sqrt{1-4 q}+1}\right)+Li_2\left(\frac{-2 \delta q+2 q+\sqrt{1-4 q}-1}{2
q+\sqrt{1-4 q}-1}\right) \right] \epsilon.
\end{eqnarray}
Note the singularity when $\delta \rightarrow 1$ in this expression.

The case of $\delta=1$ is more tricky. Here one has the overlapping of two singularities.
The argument of the hypergeometric function goes to the edge of the circle of convergence
and it is convenient to use  the integral representation
\begin{equation}
\phantom{}_2F_1(a,b,c,z) = \frac{\Gamma(c) \Gamma(b)}{ \Gamma (c-b) } \int_{0}^{1} d t
t^{b-1} (1-t)^{c - b -1} (1- t z)^{-a}.
\end{equation}
As a result, one has a two-fold integral
\begin{equation}
 \frac{\Gamma(c) \Gamma(b)}{ \Gamma (c-b) } \int_0^1 d x \ d t\ t^{b-1} (1-t)^{c - b -1}
x^{\alpha-1} (1-x)^{\beta -1}  (1- t q x^{m} (1-x)^{n})^{-a},
\end{equation}
where parameters a, b and c take the values  $a=1$, $b=- \epsilon$, $c=1 - \epsilon$.
Choosing particular values of ${\alpha,\beta,m,n}$ one can observe the overlapping
divergencies. Consider, for example, the integral
\begin{equation}  \int_0^1\!\! d x \ d t\ t^{-1-\epsilon}
x^{-1-\epsilon} (1\!-\!x)^{-2\epsilon} \frac{1}{1\!-\!qt \frac{(1\!-\!x)^2}{x}}
  = \int_0^1\!\! d x\ dt\
t^{-1-\epsilon} x^{-\epsilon} (1\!-\!x)^{-2\epsilon} \frac{1}{x\!-\!qt (1\!-\!x)^2},
\end{equation}
where we see in the last term that the denominator equals zero at $t=0$ and $x=0$. The
divergence, which occurs in this case, is the overlapping IR divergence and to handle it
we use the following trick: we insert  in the integral a unity
$$1= \Theta (x - t) + \Theta (t - x), $$
which splits the integral into two parts.  The first $\theta$-function gives
\begin{equation}   \int_0^1\ d x\  dz\ z^{-1-\epsilon} x^{-1-2\epsilon} (1-x)^{-2\epsilon}
\frac{1}{1-qz (1-x)^2},
\end{equation}
 while the other leads to
\begin{equation}  \int_0^1\ d z\ d t\ t^{-1-2\epsilon}
z^{-\epsilon} (1-zt)^{-2\epsilon} \frac{1}{z-q(1-zt)^2}.
\end{equation}

The calculation now is straightforward. One has to extract a few terms of the $\epsilon$
expansion.

For example, the first three terms of the $\epsilon$--expansion for the integral
(\ref{examp}) when $\delta=1$ are
\begin{eqnarray}
&& \int_{0}^{1} d x x^{-1-\epsilon} (1-x)^{-2 \epsilon}
\phantom{}_2F_1(1,-\epsilon;1-\epsilon; q x^{-1} (1-x)^{2})
\\ \nonumber && =-\frac{1}{\epsilon}-Li_2\left(\frac{2}{\sqrt{\frac{1}{q}-4}
\sqrt{\frac{1}{q}}-\frac{1}{q}+2}\right) \epsilon -
Li_2\left(-\frac{2}{\sqrt{\frac{1}{q}-4} \sqrt{\frac{1}{q}}+\frac{1}{q}-2}\right)
\epsilon + \frac{2 \pi ^2 \epsilon}{3} + O(\epsilon^2).
\end{eqnarray}

\section{\hspace*{-0.8cm} Appendix C. Splitting functions}
\setcounter{equation}0
\renewcommand{\theequation}{C.\arabic{equation}}

The splitting functions $P_{ij}$ which we use to calculate the splitting contribution to
the cross-section  can be obtained from the collinear limit of the color ordered tree
level partial amplitudes. Suppose one has an $n$-point partial tree amplitude in $0\leq
\mathcal{N} \leq 4$ supersymmetric gauge theory
$$A_{n}^{(tree)}(p_{a(1)}^{\lambda_{1}},...,p_{a(i)}^{\lambda_{i}},...,p_{a(n)}^{\lambda_{n}}),$$
where $a(i)$ is the color index  of $i$-th particle and $\lambda_{i}$ is it's helicity.

It can be shown~\cite{Parke,Bern1} that the MHV amplitudes have the following universal
behaviour in the collinear limit when momenta of two particles $i$ and $i+1$ become
collinear $i||i+1$
\begin{equation}\label{colin}
A_{n}^{tree}(...,p_{a}^{\lambda_{i}},p_{b}^{\lambda_{i+1}},...)\stackrel{i||i+1}
\rightarrow \ \  \sum_{\lambda,c} Split_{-\lambda}(a^{\lambda_{i}},b^{\lambda_{i+1}},z)
A_{n-1}^{tree}(...,p_{c}^{\lambda},...),
\end{equation}
where the two momenta satisfy
$$
p_{i} = z p , \ \ \  p_{i+1} = (1-z) p,
$$
 $p$ being some arbitrary momentum.  The sum goes over all possible helicities and particle
types for which $A_{n-1}^{tree}$ is nonvanishing.
The function $Split_{-\lambda}(a^{\lambda_{i}},b^{\lambda_{i+1}},z)$ depends on $p$ and
$z$. Notice the flip of helicity in
$Split_{-\lambda}(a^{\lambda_{i}},b^{\lambda_{i+1}},z)$ which comes from considering all
particles as outgoing ones.

Then the polarized version of the splitting function $P_{ij}$ can be obtained from
$Split_{-\lambda}$, up to the terms proportional to $\delta(1-z)$ by means of
\begin{equation}\label{spl}
P^{c^{-\lambda}}_{a^{\lambda_i}~b^{\lambda_{i+1}}}=
(p_i+p_{i+1})^2|Split_{-\lambda}(a^{\lambda_{i}},b^{\lambda_{i+1}},z)|^2
\end{equation}
and corresponds to the process $c\rightarrow i,i+1$ when the  particle $c$ with momentum
$p$ and helicity $\lambda$ splits into collinear particles $i$ and $i+1$ with momenta
$zp$ and $(1-z)p$ and  helicities $\lambda_i$ and $\lambda_{i+1}$, respectively.

For example, the splitting function $P^{g^-}_{g^+g^+}$ can be obtained from the partial
amplitude $A^{tree}_5(g^-g^-g^+g^+g^+)$ taking the limit $4 || 5$ ($p_4=zp$,
$p_5=(1-z)p$) in
\begin{equation}
A^{tree}_5(g^-g^-g^+g^+g^+)=
\frac{\langle12\rangle^4}{\langle12\rangle\langle23\rangle\langle34
\rangle\langle45\rangle\langle51\rangle}.
\end{equation}
One has
$$
A^{tree}_5(g^-g^-g^+g^+g^+)\stackrel{4||5} \rightarrow
\frac{1}{\langle45\rangle}\frac{1}{\sqrt{z(1-z)}}
\frac{\langle12\rangle^4}{\langle12\rangle\langle23\rangle\langle3p\rangle\langle
p1\rangle}.
$$
Thus, the only one term in the sum (\ref{colin}) survives and  $A_{n-1}^{tree}$ in this
case is $A_{4}^{tree}(g^-g^-g^+g^+)$. This gives
\begin{equation}
Split_{-}(g^+,g^+,z)=\frac{1}{\langle45\rangle}\frac{1}{\sqrt{z(1-z)}},
\end{equation}
so that, according to (\ref{spl}),
\begin{equation}
P^{g^-}_{g^+g^+}=\frac{1}{z}+\frac{1}{ (1-z)_{+}}.
\end{equation}

All the splitting functions necessary for our computation  can be obtained in a similar
fashion. They look like
\begin{eqnarray}
 P^{g^-}_{g^+g^+} &=& \frac{1}{z}+\frac{1}{ (1-z)_{+}} , \nonumber\\
  P^{g^-}_{g^+g^-} &=& \frac{z^3}{(1-z)_{+}},\nonumber\\
P^{g^-}_{g^-g^+} &=& \frac{(1-z)^3}{z}, \label{pp}\\
 P^{g^-}_{q^+\bar q^-} &=& z^2 , \nonumber\\
 P^{g^-}_{\bar{q}^-q^+} &=& (1-z)^2,  \nonumber\\
 P^{g^-}_{\Lambda\Lambda} &=& z(1-z). \nonumber
\end{eqnarray}
The contributions proportional to $\delta(1-z)$ are calculated separately from the
requirement of conservation of momenta and are absent in our case since they are
proportional to the $\beta$ function which vanishes in the $\mathcal{N}=4$ SYM theory.

The "plus" prescription in the expression $\frac{1}{(1-z)_{+}}$ in (\ref{pp}) should be
understood in the usual way:
\begin{equation}
\int_{0}^{1} d z \frac{f(z)}{(1-z)_{+}} = \int_{0}^{1} d z \frac{f(z)-f(1)}{(1-z)}.
\end{equation}
When $f(z)$ contains the theta function like in the splitting counterterm
$$
f(z) = \Theta (z - z_{min}) g(z),
$$
one has
\begin{eqnarray}&&
\int_{0}^{1} d z \frac{f(z)}{(1-z)_{+}} =\int_{0}^{1} d z \frac{\Theta (z - z_{min})
g(z)- g(1)}{(1-z)}\\&&=  \int_{z_{min}}^{1} d z \frac{g(z)- g(1)}{(1-z)} -
\int_{0}^{z_{min}} d z \frac{g(1)}{1-z} = \int_{z_{min}}^{1} d z \frac{g(z)}{(1-z)_{+}} +
\log (1-z_{min}) g(1). \nonumber
\end{eqnarray}

The splitting function $P_{qq}(z)$ (\ref{splitfunc}) used in our toy model example can be
obtained from the polarized splitting functions
\begin{eqnarray}
 P^{q^+}_{\bar q-g^-} &=& \frac{1}{(1-z)_{+}} ,\nonumber \\
 P^{q^+}_{\bar q^-g^+} &=& \frac{z^2}{(1-z)_{+}}
\end{eqnarray}
by summation over helicities. The term proportional to $\delta(1-z)$ is obtained from the
requirement of conservation of the number of quarks
$$
\int_0^1 dz\ q(z,Q_f^2/\mu^2)=1 \ \ \Rightarrow \int_0^1 dz P_{qq}(z)=0.
$$

\section{\hspace*{-0.7cm} Appendix D. Finite parts of amplitudes.}
\setcounter{equation}0
\renewcommand{\theequation}{D.\arabic{equation}}
In general all finite parts have the following structure:
$$
\mathcal{F}inite\
part=\frac{1}{(1-c^2)^2}\left[f_{Sym}(c,\delta)+(f_{Asym}(c,\delta)+f_{Asym}(-c,\delta))\right],
$$
where the functions  $f_{Sym}$ and $f_{Asym}$ contain Log\footnote{To make the
expressions more compact we use $L$ for the logarithms} and Polylog functions of $c$ and
$\delta$. Below we present the  expressions for $f_{Sym}(c,\delta)$ and
$f_{Asym}(c,\delta)$.
\renewcommand{\thesubsection}{D.\arabic{subsection}}

\subsection*{$\left(\left(\frac{d\sigma_{2\rightarrow3}}{d\Omega_{13}}
\right)^{(--+++)}_{Real}\right)_{fin}$,
general $\delta$.}
\begin{eqnarray}
f_{Sym}^{(--+++)}(c,\delta)&=& \frac{\mathcal{S}_{1}+ \mathcal{S}_{2} L(1-\delta) +
\mathcal{S}_{3} L(\delta)}{(1-\delta)^2 } -4(13+3c^2) L(\delta)L(1-\delta) \\ &&
\makebox[-7em]{} + 10(3\!+\!c^2) L^2(\delta) - 4(5\!+\!c^2)
L(\frac{1\!-\!c}{2})L(\frac{1\!+\!c}{2}) - 16(3\!+\!c^2)
L^2(1\!-\!\delta)-4(9c^2\!+\!35)Li_2(\delta),\nonumber
\end{eqnarray}
where\vspace{-0.4cm}
\begin{eqnarray}
\mathcal{S}_{1}&=& \frac43(3+c^2)\pi^2(1-\delta)^2+32(4-3\delta)\delta, \nonumber\\
\mathcal{S}_{2}&=& 4(3c^2(1-\delta)^2+37-26\delta+5\delta^2), \nonumber\\
\mathcal{S}_{3}&=& -4\delta(c^2(\delta-1)+11\delta-15) .\nonumber
\end{eqnarray}
\begin{eqnarray}
f_{Asym}^{(--+++)}(c,\delta)&=&
\frac{1}{(1-\delta)^2}\left(\mathcal{A}_{1}L(\frac{1-c}{2})
 +\mathcal{A}_{2}L(\frac{1+\delta-c(1-\delta)}{2}) \right)\nonumber \\
&-&   2(-1+4c+c^2) L^2(\frac{1-c}{2})
- 8(3+2c+c^2) L(1-\delta)L(\frac{1-c}{2})  \nonumber \\
&+& 16c L(\delta)L(\frac{1-c}{2})
+ 4(1+c)^2 L(\frac{1-c}{2})L(\frac{1+\delta-c(1-\delta)}{2}) \nonumber\\
&+& 4(1+c)^2 L(1-\delta)L(\frac{1\!+\!\delta\!-\!c(1\!-\!\delta)}{2})\! -\! 4
(1\!+\!c)^2 L(\delta)L(\frac{1\!+\!\delta\!-\!c(1\!-\!\delta)}{2}) \nonumber\\
&+&  8(1+c) Li_2(\frac{1-c}{2}) + 4(5+2c+c^2) Li_2(-\frac{\delta(1-c)}{1+c}) \nonumber\\
&-& 4(5+2c+c^2) Li_2(\frac{(1-\delta)(1-c)}{2}),
\end{eqnarray}
where\vspace{-0.3cm}
\begin{eqnarray*}
\mathcal{A}_{1}&=&4(c^2(1-\delta)^2-6c(1-\delta)^2+5+2\delta-3\delta^2),  \nonumber\\
\mathcal{A}_{2}&=&-4(c^2(1-\delta)^2-2c(3-4\delta+\delta^2)+5+2\delta-3\delta^2).
\nonumber
\end{eqnarray*}
\subsection*{$\left(\left(\frac{d\sigma_{2\rightarrow3}}{d\Omega_{13}}
\right)^{(--++-)}_{Real}\right)_{fin}$, general $\delta$.}
\begin{eqnarray}
f_{Sym}^{(--++-)}(c,\delta)&=&\mathcal{S}_{1}+\mathcal{S}_{2} L(\delta)+ \mathcal{S}_{3}
L(1\!-\!\delta)+ 10(c^2+3)
 L^2(\delta)   \\
&-&2(37+18c^2+c^4) Li_2(\delta) +\frac{8(6+9c^2+c^4)}{(1-c^2)}
L(\frac{1\!-\!c}{2})L(\frac{1\!+\!c}{2}),\nonumber
\end{eqnarray}
where
\begin{eqnarray*}
\mathcal{S}_{1}&=&\frac{8(3 c^2\!+\!5)\delta ^3 \!+\!3(7 c^2\!-\!95)\delta ^2\!+\!6(67
c^2\!+\!513)\delta}{9} +\frac 23 (c^2+3)\pi ^2+\frac{64(11c^2+7)}{3(c^2-1)}
 \nonumber \\
&& \makebox[1em]{} +\frac{32 (c^3+12 c^2+19 c+9)}{3 (1-c) (1+\delta+c(
1-\delta))}-\frac{32 (2 c^3-12 c^2+19 c-9)}{3 (1+c) (1+\delta-c(1-\delta))} \nonumber
\\ && \makebox[1em]{} -\frac{32 (c^3+4 c^2+5 c+2)}{3 (1-c) (1+\delta
+c(1-\delta))^2} +\frac{32 (c^3-4 c^2+5 c-2)}{3 (1+c) (1+\delta-c(1-\delta))^2},\nonumber\\
\mathcal{S}_{2}&=&-\frac{16}{3} (c^2+1) \delta ^3+2 (c^2+19) \delta ^2-\frac{32 (c^4+6
c^2-5) \delta }{c^2-1}+\frac{64 (12 c^2+17)}{3(c^2-1)} \nonumber
\\ && \makebox[1em]{} +\frac{32 (c^3+5 c^2+11 c+7)}{(1-c) (1+\delta+c(1-\delta))
}-\frac{32 (c^3-5 c^2+11 c-7)}{(1+c) (1+\delta -c(1-\delta))}\nonumber \\ &&
\makebox[1em]{} -\frac{32 (c^3+4 c^2+5 c+2)}{(1-c) (1+\delta+c(1-\delta))^2} +\frac{32
(c^3-4 c^2+5 c-2)}{(1+c) (1+\delta -c(1-\delta))^2}\nonumber \\ && \makebox[1em]{}
+\frac{64 (c^3+3 c^2+3 c+1)}{3 (1-c) (1+\delta+c(1-\delta))^3}-\frac{64 (c^3-3 c^2+3 c-1
)}{3 (1+c) (1+\delta-c(1-\delta))^3},  \nonumber\\
\mathcal{S}_{3}&=&-\frac{8}{3} (3 c^2\!+\!5) \delta ^3\!-\!(c^4\!+\!2 c^2\!-\!75) \delta ^2
\!-\!\frac{32 (c^4\!+\!8 c^2\!-\!11)
\delta }{c^2-1}+\frac{3 c^6\!+\!251 c^4\!+\!2953 c^2\!+\!313}{3 (c^2-1)}
\nonumber \\
&& \makebox[1em]{} -\frac{64 (c^3-5 c^2+11 c-7)}{(1+c) (1+\delta -c(1-\delta))} +\frac{64
(c^3+5 c^2+11 c+7)}{(1-c)(1+\delta +c(1-\delta))}\nonumber \\ && \makebox[1em]{}
 +\frac{64 (c^3-4 c^2+5 c-2)}{(1+c) (1+\delta -c(1-\delta))^2} -
 \frac{64 (c^3+4 c^2+5 c+2)}{(1-c)(1+\delta +c(1-\delta))^2
}\nonumber \\ && \makebox[1em]{} +\frac{128 (1+c)^3}{3(1-c) (1+\delta
+c(1-\delta))^3}+\frac{128 (1-c)^3}{3 (1+c) (1+\delta-c(1-\delta))^3}.\nonumber
\end{eqnarray*}
\begin{eqnarray}
f_{Asym}^{(--++-)}(c,\delta)&=&
\frac{8(c^4-6c^3+24c^2+6c-17)}{(1+c)^2} L(\delta)L(\frac{1-c}{2}) \nonumber\\
&+&\frac{4(3+c^2)^2}{(1+c)^2} L(\frac{1-c}{2})L(\frac{1+\delta-c(1-\delta)}{2})
\\
&-&\frac{4(7+c^2)(1-c)}{(1+c)}L(\frac{1+c}{2})
L(\frac{1+\delta-c(1-\delta)}{2}) \nonumber\\
&-&\frac{8(c^4-12c^3+34c^2+12c-43)}{(1+c)^2} L(1-\delta)L(\frac{1+\delta-c(1-\delta)}{2})\nonumber\\
&-&\frac{2(c^4-2c^3+8c^2-6c+15)}{(1+c)^2}L^2(\frac{1-c}{2})  \nonumber\\
&+&\mathcal{A}_{1}L(\frac{1-c}{2})+\mathcal{A}_{2}L(\frac{1+\delta-c(1-\delta)}{2})
\nonumber\\ &-& \frac{4 \left(3 c^4-12 c^3+46 c^2+12 c-33\right)}{(1+c)^2}
L(\delta)L(\frac{1+\delta-c(1-\delta)}{2})\nonumber\\
&+&\frac{(c^6-2c^5+3c^4-76c^3-153c^2+14c+149)}{(1-c)^2}
Li_2(-\frac{1\!-\!c}{1\!+\!c}\delta)  \nonumber\\ &+&
\frac{8(c^4+12c^3+34c^2-12c-43)}{(1-c)^2} \left( Li_2(\frac{1\!-\!c}{2}) -
Li_2(\frac{(1\!-\!\delta)(1\!-\!c)}{2}) \right),\nonumber
\end{eqnarray}
where\vspace{-0.2cm}
\begin{eqnarray}
\mathcal{A}_{1}&=&-\frac{1}{6(1+c)} \left(
2597+240\delta-105\delta^2+24\delta^3+3c^5(\delta^2-1)+3c^4(3+\delta^2) \right. \nonumber
\\  && \left.
-2c^3(111-24\delta+9\delta^2-4\delta^3)+2c^2(489\!+\!120\delta\!-\!69\delta^2\!+\!20
\delta^3)\right. \nonumber  \\
&&\left. -c(2655\!-\!240\delta\!+\!225\delta^2\!-\!56\delta^3) \right) ,\nonumber \\
\mathcal{A}_{2}&=&\frac{1}{6(1\!+\!c)} \left(3 (\delta^2-1) c^5+3 (\delta^2+3) c^4+2 (8
\delta^3-3 \delta^2+36 \delta-111) c^3 \right. \nonumber \\ && \left.  +6 (8 \delta^3-17
\delta^2+28 \delta+163) c^2+3 (16 \delta^3-63 \delta^2+104 \delta-885) c \right.
\nonumber\\ && \left.+16 \delta^3-93 \delta^2+216 \delta+2597 \right).\nonumber
\end{eqnarray}
In the case $\delta=1$ one gets major simplifications:
\begin{eqnarray}
f_{Sym}^{(--++-)}(c,1) &=& \frac{2257-93\pi^2-3c^4\pi^2+c^2(303-48\pi^2)}{9} \nonumber \\
&+&\frac{8(6+9c^2+c^4)}{1-c^2} L(\frac{1-c}{2})L(\frac{1+c}{2}),
\end{eqnarray}
\begin{equation}
f_{Asym}^{(--++-)}(c,1)=\frac{\mathcal{A}_{1}}{1+c}L(\frac{1-c}{2})+
\frac{\mathcal{A}_{2}}{(1+c)^2}L^2(\frac{1-c}{2}) +
\frac{\mathcal{A}_{3}}{1-c}Li_2(\frac{1-c}{2})
\end{equation}
where\vspace{-0.4cm}
\begin{eqnarray}
\mathcal{A}_{1}&=&-\frac23 (3c^4-46c^3+280c^2-646c+689),\nonumber\\
\mathcal{A}_{2}&=&-\frac 12(c^6+2 c^5+7 c^4+68 c^3-121 c^2-38 c+209) , \nonumber\\
\mathcal{A}_{3}&=&c^5-c^4-6 c^3-178 c^2-603 c-493.\nonumber
\end{eqnarray}
\subsection*{$\left(\left(\frac{d\sigma_{2\rightarrow3}}{d\Omega_{13}}\right)^{(--+q
\bar{q})}_{Real}\right)_{fin}$, $\delta=1$}
\begin{equation}
f_{Sym}^{(--+q \bar{q})}(c,1)= - \frac{32 \left(c^2\!+\!1\right)^2}{(1\!-\!c^2)}
L(\frac{1-c}{2})L(\frac{1+c}{2}) - \frac{4 \pi ^2 (1-c^4)+132(c^2+3)}{3},
\end{equation}
\begin{equation}
f_{Asym}^{(--+q \bar{q})}(c,1)= \frac{\mathcal{A}_1}{(1+c)} L(\frac{1-c}{2}) +
\frac{\mathcal{A}_2}{(1+c)} L^2(\frac{1-c}{2}) + \frac{\mathcal{A}_3}{(1-c)}
Li_2(\frac{1-c}{2}),
\end{equation}
where\vspace{-0.1cm}
\begin{eqnarray*}
\mathcal{A}_1&=& \frac{8}{3} (3 c^4-44 c^3+222 c^2-450 c+277),\\
\mathcal{A}_2&=&-2(c^4+2 c^3-2 c^2+50 c-67) (1-c),\\
\mathcal{A}_3&=&- 4 (c^4-2 c^3-18 c^2-146 c-211) (1+c).
\end{eqnarray*}
\subsection*{$\left(\left(\frac{d\sigma_{2\rightarrow3}}{d\Omega_{13}}
\right)^{(--+ \Lambda\Lambda)}_{Real}\right)_{fin}$, $\delta=1$}
\begin{equation}
f_{Sym}^{(--+ \Lambda \Lambda)}(c,1)=-24 (c^2\!+\!1)
L(\frac{1\!-\!c}{2})L(\frac{1\!+\!c}{2}) + 6 (11 c^2\!-\!3) - \pi^2 (1\!-\!c^2)^2,
\end{equation}
\begin{eqnarray}
f_{Asym}^{(--+ \Lambda \Lambda)}(c,1)&=& -\frac{3
(c+5)\left(c^2-2 c+9\right)(1-c)^2}{2(1+c)} L^2(\frac{1-c}{2}) \\
\nonumber &-&\frac{2 \left(3c^4-47 c^3+213 c^2-369 c+184 \right)}{1+c} L(\frac{1-c}{2})
\\ \nonumber &+&\frac{3 \left(c^3-3 c^2-17c-125\right)(1+c)^2}{(1-c)}
Li_2(\frac{1-c}{2}).
\end{eqnarray}
\subsection*{$\left(\left(\frac{d\sigma_{2 \rightarrow
3}}{d\Omega_{13}}\right)^{(--+++)}_{InSplit}\right)_{fin}$, general $\delta$.}
\begin{equation}
f_{Sym}^{(--+++)}(c,\delta)=16 \frac{\delta (3 \delta-
4)}{(1-\delta)^2}+\frac{32(\delta-2)}{(1-\delta)^2}L(1-\delta)+8(3+c^2)L^2(1-\delta),
\end{equation}
\begin{eqnarray}
f_{Asym}^{(--+++)}(c,\delta)&=& 4 (c^2+3) \left(  L^2(\frac{1-c}{2}) -2 L(\frac{1-c}{2})L(\frac{1+\delta-c(1-\delta)}{2}) \right. \nonumber\\
&+& L^2(\frac{1+\delta-c(1-\delta)}{2}) + Li_2(\frac{1-c}{2}) - 2Li_2(-\frac{(1-c)(1-\delta)}{2})\nonumber\\
&+& 2\left. Li_2(-\frac{(1-c)\delta}{1+c})+ 2
Li_2(\frac{2\delta}{1+\delta-c(1-\delta)})\right).
\end{eqnarray}
\subsection*{$\left(\left(\frac{d\sigma_{2\rightarrow3}}{d\Omega_{13}}
\right)^{(--++-)}_{InSplit}\right)_{fin}$, general $\delta$.}
\begin{equation}
f_{Sym}^{(--++-)}(c,\delta)=\mathcal{S}_1L(1-\delta)+\mathcal{S}_2,
\end{equation}
where
\begin{eqnarray}
\mathcal{S}_1&=&\frac{16}{3} (c^2+1) \delta ^3+8 (c^2-5) \delta ^2+16 (c^2+17) \delta
-\frac{8 (27 c^4+378 c^2+59)}{3 (c^2-1)} \nonumber
\\ &&\hspace*{-0.5cm}  -\frac{64 (c^3+5 c^2+11 c+7)}{(1-c) (1+\delta +c(1-\delta))}+\frac{64(c^3-5 c^2
+11 c-7)}{(1+c) (1+\delta -c(1-\delta))}+\frac{64 (c^3+4 c^2+5 c+2)}{(1-c) (1+\delta
+c(1-\delta)^2} \nonumber
\\ &&\hspace*{-0.5cm}  -\frac{64 (c^3-4 c^2+5 c-2)}{(1+c)
(1+\delta -c(1-\delta))^2}-\frac{128 (c^3+3 c^2+3 c+1)}{3 (1-c) (1\!+\!\delta\!+\!
c(1\!-\!\delta))^3}+\frac{128 (c^3-3 c^2+3 c-1)}{3
(1+c) (1\!+\!\delta \!-\!c(1\!-\!\delta))^3},\nonumber \\
\mathcal{S}_2&=& -\frac{16}{9} (c^2+1) \delta ^3-\frac{4}{3}(5 c^2-13) \delta
^2-\frac{8}{3} (11 c^2+89) \delta -\frac{256(2 c^2+1)}{3 (c^2-1)} \nonumber
\\ && \makebox[1em]{} -\frac{32 (2 c^3+9 c^2+12 c+5)}{3
(1-c) (1+\delta +c(1-\delta))}+\frac{32 (2 c^3-9 c^2+12 c-5)}{3 (1+c) (1+\delta
-c(1-\delta))} \nonumber
\\ && \makebox[1em]{} +\frac{32 (c^3+3
c^2+3 c+1)}{3 (1-c) (1+\delta +c(1-\delta))^2}-\frac{32 (c^3-3 c^2+3 c-1)}{3 (1+c)
(1+\delta -c(1-\delta))^2}.
\end{eqnarray}
\begin{eqnarray}
f_{Asym}^{(--++-)}(c,\delta)&=&\frac{16(1-c)(4c^2-17c+37)}{3(1+c)} \left(
L(\frac{1\!+\!\delta\!-\!c(1\!-\!\delta)}{2}) - L(\frac{1-c}{2}) \right) \nonumber\\&+&
4(3\!+\!c^2) \left( L(\frac{1\!+\!\delta\!-\!c(1\!-\!\delta)}{2}) - L(\frac{1-c}{2} )
\right)^2 \nonumber \\ &+& \frac{8 (c^3-15 c^2+51 c-45)}{1+c} L(1-\delta)
L(\frac{1\!+\!\delta\!-\!c(1\!-\!\delta)}{2})
\\ &+&8(3+c^2) \left( Li_2(-\frac{1-c}{1+c}\delta)-
Li_2(\frac{2\delta}{1+\delta-c(1-\delta)}) \right) \nonumber\\
&+&\frac{8 \left(c^3+15 c^2+51 c+45\right)}{1-c} \left( Li_2(\frac{1-c}{2}) - Li_2(\frac
12 (1-c)(1-\delta) ) \right).\nonumber
\end{eqnarray}
In the case $\delta=1$ one gets major simplifications:
\begin{equation}
f_{Sym}^{(--++-)}(c,1)=\frac 49 \left( (6 \pi^2-49) c^2+18 \pi^2 - 415 \right),
\end{equation}
\begin{equation}
f_{Asym}^{(--++-)}(c,1)= \frac{16(1\!-\!c)\left(4 c^2\!-\!17 c\!+\!37\right) }{3(1+c)}
L\!\left(\!\frac{1\!-\!c}{2}\!\right) + \frac{16(1\!+\!c)\left(c^2\!+\!6
c\!+\!21\right)}{1-c} Li_2\!\left(\!\frac{1\!-\!c}{2}\!\right).
\end{equation}\vspace{-0.8cm}

\subsection*{$\left(\left(\frac{d\sigma_{2\rightarrow3}}{d\Omega_{13}}\right)^{(--+q
\bar{q})}_{InSplit}\right)_{fin}$, $\delta=1$}
\begin{equation}
f_{Sym}^{(--+q \bar{q})}(c,1)=\frac{16}{3} (9 c^2+23),
\end{equation}
\begin{equation}
f_{Asym}^{(--+q \bar{q})}(c,1)= -\frac{64 (4 c^2-17 c+19)(1-c)}{3(1+c)} L(\frac{1-c}{2})
-\frac{64(c+3)^2 (1+c)}{(1-c)} Li_2(\frac{1-c}{2}).
\end{equation}

\subsection*{$\left(\left(\frac{d\sigma_{2\rightarrow3}}{d\Omega_{13}}
\right)^{(--+ \Lambda\Lambda)}_{InSplit}\right)_{fin}$, $\delta=1$}
\begin{equation}
f_{Sym}^{(--+ \Lambda \Lambda)}(c,1)= -16 (3 c^2-1),
\end{equation}
\begin{equation}
f_{Asym}^{(--+ \Lambda \Lambda)}(c,1)= \frac{16 (13-4c)(1-c)^2}{(1+c)} L(\frac{1-c}{2}) +
\frac{48 (c+5)(1+c)^2}{(1-c)} Li_2(\frac{1-c}{2}).
\end{equation}

\end{document}